\documentclass[twocolumn,prb,showpacs,preprintnumbers,amsmath,amssymb,superscriptaddress]{revtex4}
\usepackage{graphicx}
\usepackage{setspace}

\usepackage{color}
\usepackage{ulem} 

\usepackage{amsmath,amsthm,amssymb}
\usepackage{mathrsfs}

\newcommand{\sigz}{\hat{\sigma}_3}

\makeatletter
\def\Left#1#2\Right{\begingroup%
   \def\ts@r{\nulldelimiterspace=0pt \mathsurround=0pt}%
   \let\@hat=#1%
   \def\sht@im{#2}%
   \def\@t{{\mathchoice{\def\@fen{\displaystyle}\k@fel}%
          {\def\@fen{\textstyle}\k@fel}%
          {\def\@fen{\scriptstyle}\k@fel}%
          {\def\@fen{\scriptscriptstyle}\k@fel}}}%
   \def\g@rin{\ts@r\left\@hat\vphantom{\sht@im}\right.}%
   \def\k@fel{\setbox0=\hbox{$\@fen\g@rin$}\hbox{%
      $\@fen \kern.3875\wd0 \copy0 \kern-.3875\wd0%
      \llap{\copy0}\kern.3875\wd0$}}%
      \def\pt@h{\mathopen\@t}\pt@h\sht@im%
      \Right}%
\def\Right#1{\let\@hat=#1%
   \def\st@m{\mathclose\@t}%
   \st@m\endgroup}
\makeatother

\pacs{71.10.Fd}

\begin{document}
\title{Nonequilibrium steady states and transient dynamics of conventional superconductors under phonon driving
}

\author{Yuta Murakami}
\affiliation{Department of Physics, University of Fribourg, 1700 Fribourg, Switzerland}
\author{Naoto Tsuji}
\affiliation{RIKEN Center for Emergent Matter Science (CEMS), Wako 351-0198, Japan}
\author{Martin Eckstein}
\affiliation{Max Planck Research Department for Structural Dynamics, University of Hamburg-CFEL, 22761 Hamburg, Germany}
\author{Philipp Werner}
\affiliation{Department of Physics, University of Fribourg, 1700 Fribourg, Switzerland}
\date{\today}

\begin{abstract}
We perform a systematic analysis of the influence of phonon driving on the superconducting Holstein model coupled to heat baths
by studying both the transient dynamics and the nonequilibrium steady state (NESS) in the weak and strong electron-phonon coupling regimes. Our study is based on the nonequilibrium dynamical mean-field theory, and for the NESS we present a Floquet formulation 
adapted to electron-phonon systems. 
The analysis of the phonon propagator suggests that the effective attractive interaction can be strongly enhanced in a parametric resonant regime
because of the Floquet side bands of phonons.
While this may be expected to enhance the superconductivity (SC), our fully self-consistent calculations, which include the  
effects of heating and nonthermal distributions, show that the parametric phonon driving generically results in a suppression or complete melting of the SC order. 
In the strong coupling regime, the NESS always shows a suppression of the SC gap, the SC order parameter and the superfluid density as a result of the 
driving, and 
this tendency is most prominent at the parametric resonance. 
Using the real-time nonequilibrium DMFT formalism, we also study the dynamics towards the NESS, which shows that the heating effect dominates the transient dynamics, and SC is weakened by the external modulations, in particular at the parametric resonance.
In the weak coupling regime, we find that the SC fluctuations above the transition temperature are generally weakened under the 
driving. The strongest 
suppression occurs again around the parametric resonances because of the efficient energy absorption.
\end{abstract}

\maketitle

\section{Introduction}
The prospect of nonequilibrium exploration and control of material properties has caught the interest and attention of broad segments of the condensed matter community. 
\cite{Giannetti2016}
In particular, the resonant excitation of mid-infrared phonon modes has opened a new pathway for manipulating  
properties such as metallicity,\cite{Cavalleri2007} magnetism\cite{Forst2011b} and  superconductivity (SC),
\cite{Cavalleri2011,Kaiser2014,Hu2014,Mankowsky2014,Mitrano2016,Cantaluppi2017}  and for 
modifying lattice structures in ways which may not be achievable in equilibrium.
Especially, the light-induced SC-like behavior in the optical conductivity for high $T_c$ cuprates and doped fullerides
\cite{Cavalleri2011,Kaiser2014,Hu2014,Mankowsky2014,Mitrano2016} has stimulated several theoretical investigations.\cite{Sentef2015,Knap2016,Komnik2016,Kennes2017,Kim2016,Okamoto2016,Sentef2017,Babadi2017,Mazza2017,Fernandes2017}
So far, two possible scenarios for the enhancement of SC have been discussed in the literature based on the enhanced pairing interaction out of equilibrium: 
(i) Under time-periodic driving, the time-averaged Hamiltonian can be modified in such a way that superconductivity is favored, \cite{Sentef2015,Kim2016,Mazza2017} and (ii) the interaction can be effectively enhanced in the driven state by dynamical effects. \cite{Knap2016,Komnik2016,Kennes2017,Sentef2017}
Both effects can be induced by the resonant excitation of mid-infrared phonon modes, and previous theoretical investigations have tried to clarify their influence on SC.

Concerning the effect of time-periodic driving on the time-averaged Hamiltonian, 
it has been pointed out that the resonantly excited mid-infrared phonon can lead to favorable parameter changes,\cite{Sentef2015,Kim2016,Mazza2017} such as an orbital imbalance of the Coulomb interaction.
As for the effects of the dynamical part of the Hamiltonian, it has been argued that
the parametric driving of the Raman modes through nonlinear couplings with the mid-infrared phonon modes\cite{Knap2016,Komnik2016} 
or  the excited mid-infrared phonon through a nonlinear electron-phonon coupling\cite{Kennes2017,Sentef2017} 
effectively pushes the system into the stronger-coupling 
regime, leading to an enhancement of the SC.
Still, many important questions remain to be answered. 
In particular, when the enhancement of the SC is attributed to a change of the time-averaged Hamiltonian or the effective interaction, it is often assumed that the system remains in thermal equilibrium. 
However, in general, a periodically driven and interacting many-body system is expected to heat up (either by resonant excitations, or on a longer timescale by multi-photon absorption processes) and eventually reach an infinite temperature state.\cite{Rigol2014,Ponte2015}
This would imply a melting of long-range order after a long time, even if the pairing interactions are enhanced by the driving. 
In a real system, this detrimental heating effect may potentially be circumvented in two ways: Either the heating processes are slow enough so that an enhancement of a long-range order can be observed at least transiently, or the heating is eventually balanced by energy dissipation to the environment, thus leading to the so-called nonequilibrium steady state (NESS). The transient dynamics has been studied by looking at the SC instabilities of the normal phase, in the form of negative eigenvalues of the transient pairing susceptibility.\cite{Knap2016,Babadi2017}
While such negative susceptibilities can occur for short times,
it is an open question on which time scale the long-range order could evolve in such a situation,
and whether a potential increase can be fast enough to overcome the heating. Even less is known about the nature of the NESSs, where excitation and dissipation are balanced. 
For an understanding of the light-enhanced SC, it is therefore essential to go beyond the study of effective interactions and Hamiltonians, and to consider a formalism that can take account of the transient or steady-state modifications of the electron and phonon distribution functions, and of the SC order or fluctuations,  in the nonequilibrium state.

In this paper, we focus on the dynamical aspect of a phonon-driven system and discuss the influence of the parametric phonon driving 
and the resulting heating and nonthermal distribution on the SC.
Specifically, we study the time-periodic nonequilibrium steady state (NESS) and the nonequilibrium transient dynamics of the Holstein model, a prototypical model of electron-phonon systems, using the nonequilibrium dynamical mean-field theory (DMFT) combined with the Migdal approximation.\cite{Aoki2013,Murakami2015,Murakami2016,Murakami2016b}
To investigate the time-periodic NESS of the electron-phonon (el-ph) system with externally driven phonons,  
we extend the Floquet DMFT formalism, which has been previously only applied to purely electronic systems,\cite{Schmidt2002,Joura2008,Tsuji2008,Tsuji2009,Lee2014,Mikami2016} to the Holstein case.  

The Floquet DMFT is based on the Floquet Green's function formalism, which can naturally take into account the effect of many-body interactions, periodic driving by external fields and a dissipative coupling to the environment on equal footing. 
The Floquet DMFT for el-ph systems developed here is therefore not limited to the SC order, and should be useful to study a wide range of nonequilibrium situations that involve phonon excitations triggered by ``nonlinear phononics".\cite{Forst2011a}

Our systematic analysis, which covers the weak and strong coupling regimes, reveals that 
the parametric phonon driving generally suppresses the SC (see Fig.~\ref{fig:FDMFT_summary_dum} and Fig.~\ref{fig:sc_fluc_summary}), 
even though in some parameter regimes the retarded attractive interaction is increased and an enhancement of the SC is naively expected (see Fig.~\ref{fig:D0_info}(a)(d)).
In particular in the parametrically resonant regime, where the system absorbs energy most efficiently independent of the coupling strength, 
the SC order and fluctuations are strongly suppressed.

The paper is organized as follows.
Section~\ref{sec:setup} describes the setup for the Holstein model under external phonon driving and attached to thermal baths.
The framework used to study the problem is explained in Sec.~\ref{sec:theory} and the nonequilibrium Green's functions are introduced in Sec.~\ref{sec:noneq_form}. 
In Sec.~\ref{sec:thermal_bath}, we discuss the heat bath directly coupled to the phonon degrees of freedom, and in Sec.~\ref{sec:GF} 
we introduce the Floquet Green's function and derive its expression for the case of parametric phonon driving.
In Sec.~\ref{sec:FDMFT}, we explain the Floquet DMFT for the Holstein model and provide expressions for relevant physical 
quantities.
Important remarks about the usage and limitations of the Floquet DMFT for the parametric phonon driving problem can be found in Sec.~\ref{sec:para_ins}.
The results of our systematic analysis are shown in Sec.~\ref{sec:results}.
In Sec.~\ref{sec:D0}, we focus on the modification of the phonon propagator in the driven state in the weak coupling limit, and show that an enhancement of the attractive interaction is realized in some parameter regimes,
which naively suggests 
an enhancement of SC.
Sections~\ref{sec:NESS} and~\ref{sec:Trans1} are dedicated to the strong coupling regime.
In Sec.~\ref{sec:NESS}, we study the NESS under the parametric phonon driving using the Floquet DMFT and in Sec.~\ref{sec:Trans1} we discuss the transient dynamics toward the NESS.
In Sec.~\ref{sec:Trans2}, we focus on the weak coupling regime, and study how the SC fluctuations above $T_c$ react to the phonon modulation. 
A summary and conclusions are provided in Sec.~\ref{sec:conclusion}

\section{Model}\label{sec:setup}
\begin{figure}[t]
  \centering
   \includegraphics[width=70mm]{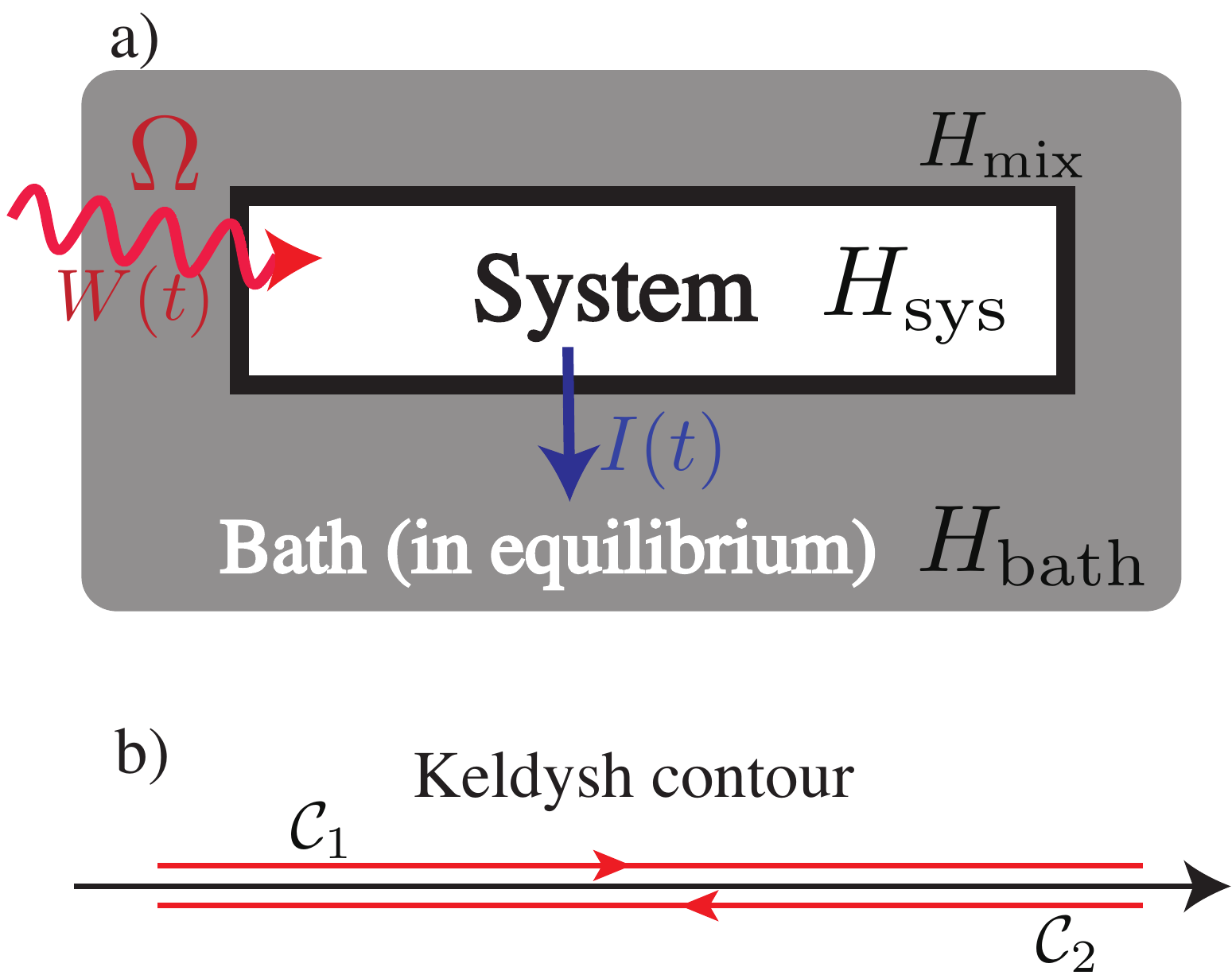}
  \caption{ (a) Schematic picture of the setup we consider in the paper. $\Omega$ is the frequency of the external field, $W(t)$ is the energy injected from the field, while $I(t)$ is the energy flow from the system to the bath. (b) The Keldysh contour.}
  \label{fig:setup}
\end{figure}

In this paper, we consider the case where the system coupled to heat baths 
is periodically driven by an external field.
Such a situation is expressed by the Hamiltonian
\begin{align}
H_{\rm tot}(t)=H_{\rm sys}(t)+H_{\rm mix}+H_{\rm bath},
\end{align}
see Fig.~\ref{fig:setup}(a).
Here $H_{\rm sys}(t)$ is the Hamiltonian for the system, which includes terms representing 
periodic excitations.
 $H_{\rm bath}$ is the Hamiltonian for heat baths,
while $H_{\rm mix}$ represents the coupling between the system and the baths.
When the system is continuously excited by periodic fields, it is expected that 
it finally reaches a time-periodic nonequilibrium steady state (NESS) in which the energy injection from the field and the energy flow into the baths are balanced.
This situation is different from isolated systems under periodic driving.\cite{Rigol2014,Ponte2015,Kuwahara2016,Mori2016}

The system Hamiltonian $H_{\rm sys}(t)$ consists of the unperturbed part $H_{\rm sys, 0}$ and the external periodic field, $H_{\rm ext}(t)$.
Here, we focus on a simple but relevant example of an electron-phonon coupled system, the Holstein model, with the Hamiltonian 
\begin{align}
H_{\rm sys,0}
=&-v\sum_{\langle  i,j\rangle,\sigma}(c_{i,\sigma}^{\dagger}c_{j,\sigma}+{\rm {\rm h.c.}})-\mu\sum_i n_i\nonumber\\
&+\omega_0\sum_i a^{\dagger}_i a_i+g\sum_i (a_i^{\dagger}+a_i)(n_i-1),\label{eq:Holstein}
\end{align}
where $c_i^\dagger$ is the creation operator for an electron with spin $\sigma$ at site $i$, 
$v$ is the electron hopping, $\mu$ is the 
electron chemical potential, $n_i = 
c_{i,\uparrow}^\dagger c_{i,\uparrow}+
c_{i,\downarrow}^\dagger c_{i,\downarrow}$, 
$\omega_0$ is the bare phonon frequency, $a_i^\dagger$ is the creation operator for the Einstein phonon, 
and $g$ is the el-ph coupling. 
The retarded attractive interaction between electrons is mediated by the phonons, which leads to 
an s-wave superconducting state at low enough temperatures. 
The SC order parameter is defined as $\phi=\frac{1}{N}\sum_i \langle c_{i\downarrow} c_{i\uparrow}\rangle$ which is assumed to be real without loss of generality.
The strength of the phonon-mediated attractive interaction is characterized by the dimensionless el-ph coupling, $\lambda=-\rho_0(0)g^2 D^R(\omega=0)$, where 
$\rho_0(\omega)$ is the density of states for the free electrons, 
and $D$ is the retarded phonon propagator.%
\cite{Bauer2011}

Recent developments in THz or mid-infrared laser techniques make it possible to selectively and strongly excite a certain phonon mode.
The infrared active phonons can couple to the electron sector through further excitation of another phonon mode (Raman mode) that is linearly 
coupled to the electron sector\cite{Subedi2014,Knap2016} or through 
a nonlinear-coupling mechanism.\cite{Mitrano2016,Singla2015,Kennes2017}
Our present interest is in the former case. 
Considering the mid-infrared phonon oscillations as an external driving force to the Raman mode through the nonlinear coupling,
we study the following 
type of excitations for the phonon sector:
\begin{align}
H_{\rm ph}(t)=\omega_{0P}(t)\sum_i \frac{P_i^2}{4}+\omega_{0X}(t)\sum_i \frac{X_i^2}{4}.\label{eq:excitation}
\end{align}
Here $H_{\rm ph}(t)$ consists of the time-dependent part ($H_{\rm ext}(t)$) and the time-independent part ($\omega_0\sum_i a^{\dagger}_i a_i$ in $H_{\rm sys,0}$).
$\omega_{0P}(t)$ and $\omega_{0X}(t)$ are periodic in time with a period $\mathscr{T}\equiv\frac{2\pi}{\Omega}$, 
$X_i=a_i+a_i^\dagger$ and $P_i=(a_i-a_i^\dagger)/i$.
This type of modulation of the phonon degrees of freedom has been discussed as a potential origin of the enhancement of superconductivity.\cite{Knap2016,Komnik2016,Babadi2017}

To be specific, we focus on two types of periodic phonon excitations,
\begin{align}
\text{ Type 1 : }&{H_{\rm ph}(t)}=\omega_{0}\sum_i\frac{P_i^2}{4}+\omega_{0X}(t)\sum_i\frac{X_i^2}{4}\label{eq:type1}\\
&\text{with    } \omega_{0X}(t)=\omega_0+\Delta\omega_0 \cos(\Omega t),\nonumber
\end{align}
and 
\begin{align}
\text{ Type 2 : }&H_{\rm ph}(t)=\omega_{0}(t) \sum_i\left(\frac{P_i^2}{4}+\frac{X_i^2}{4}\right)\label{eq:Komnik}\\
&\text{with    } \omega_{0}(t)=\omega_0+\Delta\omega_0 \cos(\Omega t).\nonumber
\end{align}
The first type, Eq.~(\ref{eq:type1}), is included in excitations through some of the nonlinear couplings between the mid-infrared mode and the Raman mode.
Its potential effect on superconductivity has been discussed in Ref.~\onlinecite{Knap2016}.
Potential effects of the second type of excitations, Eq.~(\ref{eq:Komnik}), have been discussed in Ref.~\onlinecite{Komnik2016}.
As we will show below, it turns out that the type 1 and type 2 excitations give qualitatively the same results concerning the enhancement of the attractive interaction and the 
suppression of SC in the NESS and the transient dynamics.
Besides these two types of excitations, one may also want to consider the case where $\frac{1}{\mathscr{T}}\int^\mathscr{T}_0\omega_{0X}(t)dt \neq\omega_0$ as in Ref.~\onlinecite{Knap2016}. 
If $\frac{1}{\mathscr{T}}\int^\mathscr{T}_0\omega_{0X}(t)dt<\omega_0$, the system goes to the stronger-coupling regime and the SC should be enhanced.
However, since this effect is rather trivial we do not discuss it in this paper.

 As for the bath, we use free electron and free phonon baths, which are introduced in detail in the 
following section. 

\section{Formalism}\label{sec:theory}

\subsection{Nonequilibrium Green's functions}\label{sec:noneq_form}

The transient time evolution toward the NESS can be described in the nonequilibrium Green's function method formulated on the 
Kadanoff-Baym (KB) contour $\mathcal C_{\rm KB}$,\cite{Aoki2013} where we define the contour-ordered Green's functions,
\begin{subequations}
\begin{align}
G(t,t')&\equiv -i \langle \mathcal{T}_{\mathcal {C}_{\rm KB}} c(t)c^\dagger(t')\rangle, \\
D(t,t')&\equiv -i \langle \mathcal{T}_{\mathcal {C}_{\rm KB}} X(t)X(t')\rangle, 
\end{align}
\end{subequations}
for the electrons and phonons, respectively. 
Here $c^\dagger$ is the creation operator for an electron in the system, $X$ represents the phonon distortion 
and $\mathcal{T}_{\mathcal {C}_{\rm KB}}$ is the contour-ordering operator on the KB contour. All the operators are in the Heisenberg picture,
and $\langle \cdots \rangle$ indicates the average with respect to the initial ensemble.  
In this formalism, the system is in equilibrium at the initial time $t_0$ and is then driven by the periodic external field.
In Secs.~\ref{sec:Trans1},\ref{sec:Trans2} we use this formalism to study the transient dynamics. 
The DMFT formalism for the real-time dynamics, both in electron and electron-phonon-coupled systems, has been discussed in previous works\cite{Aoki2013,Werner2013,Murakami2015,Murakami2016} and will not be presented in detail here.

The KB formalism can be inefficient for studying the NESS, which the system reaches in the long-time limit. 
Since the initial correlations are wiped out through the coupling to the thermal bath, they should be irrelevant 
to the NESS.\cite{Aoki2013} Hence,  we can simplify the problem by assuming $t_0=-\infty$ and neglecting the left-mixing, right-mixing and Matsubara components
in the KB formalism (which reduces to the Keldysh formalism), and focus directly on the NESS.
\footnote{We assume that at $t_0=-\infty$ the system is free and decoupled from the baths, and that the interactions and the coupling to the baths are adiabatically turned on.}
In the Keldysh formalism, we focus on the contour $\mathcal{C}_K=\mathcal{C}_1\cup \mathcal{C}_2$ [see Fig.~\ref{fig:setup}(b)], 
and hence only need to consider 
\begin{align}
\check{G}(t,t')\equiv
\begin{bmatrix}
G^{11}(t,t')&G^{12}(t,t')\\
G^{21}(t,t')&G^{22}(t,t')
\end{bmatrix}.
\end{align}
Here $G^{ij}(t,t')$ indicates $t\in \mathcal{C}_i$ and $t'\in \mathcal{C}_j$.

The physical representation for the Green's function is defined  
as 
\begin{align}
\underline{G}&\equiv
\begin{bmatrix}
G^{R}&G^K\\
0&G^A
\end{bmatrix}=\check{L}\check{\sigma}_3\check{G}\check{L}^\dagger
\text{ with } 
\check{L}=\frac{1}{\sqrt{2}}
\begin{bmatrix}
1&-1\\
1&1
\end{bmatrix},
\end{align}
where $\check{\sigma}_{3}$ is the third component of the Pauli matrices.
This is the so-called Larkin-Ovchinnikov form.
We can apply the same transformation to the phonon Green's function $D$.

If there is no interaction or coupling to external baths, 
one can evaluate the bare Green's function $G_0$ for electrons ($D_0$ for phonons), even in the presence of an external field. 
The correlations coming from the interaction and baths can be taken into account through the self-energy $\Sigma$ for electrons ($\Pi$ for phonons).
Since the initial correlation can be neglected,
the Dyson equation in the Keldysh formalism becomes simple:
\begin{subequations}\label{eq:Dyson_time}
\begin{align}
\int^\infty_{-\infty} dt_1[\underline{G}_0^{-1} (t,t_1)-\underline{\Sigma}(t,t_1)] \cdot \underline{G}(t_1,t')=\underline{I}\delta(t-t'),\label{eq:Dyson_Keldysh}\\
\int^\infty_{-\infty} dt_1[\underline{D}_0^{-1} (t,t_1)-\underline{\Pi}(t,t_1)] \cdot \underline{D}(t_1,t')=\underline{I}\delta(t-t').\label{eq:ph_dyson_time}
\end{align}
\end{subequations}
Here $\cdot$ indicates the matrix product, $\underline{I}$ is the $2\times2$ identity matrix, $\int dt_1\underline{G}_0^{-1} (t,t_1) \underline{G}_0(t_1,t')=\underline{I}\delta(t-t')$ (the same holds for $\underline{D}_0^{-1}$).
Precisely speaking, Eq.~(\ref{eq:ph_dyson_time}) is true as long as the average phonon distortion $\langle X(t)\rangle$ remains zero, which is the case considered in this paper.
When $\langle X(t)\rangle$ is finite, the phonon propagator has an additional contribution $-i\langle X(t)\rangle \langle X(t')\rangle$.

\subsection{Thermal baths}\label{sec:thermal_bath}
In this paper, we consider two types of thermal baths. 
One is attached to the electron degrees of freedom and the other is attached to the phonon degrees of freedom. 

For the bath attached to the electrons we employ the so-called B\"{u}ttiker model,
which consists of multiple sets of free electrons in equilibrium connected to each site of the system.
The model has been used in the previous Floquet DMFT studies for electron systems.\cite{Tsuji2009,Mikami2016, Aoki2013}
Here we briefly review the model. Due to the noninteracting nature of the bath, it can be traced out exactly to yield the self-energy correction $\Sigma_{{\rm bath}}$,
which is characterized by 
\begin{align}
\Gamma(\omega)\equiv-{\rm Im}\Sigma^R_{\rm bath}(\omega).
\end{align}
All of the physical components of $\Sigma_{{\rm bath}}$ can be obtained through the fluctuation-dissipation theorem for 
$\Sigma_{\rm bath}(\omega)$.\cite{Aoki2013}
The total self-energy is the sum of the contributions from the bath and the interactions in $H_{\rm sys}$,
\begin{align}
\Sigma_{{\rm tot}}=\Sigma_{{\rm bath}}+\Sigma_{{\rm int}}.
\end{align}

In the following, we use a heat bath with a finite band width,
\begin{align}
\Gamma(\omega)=\gamma_{\text {el}} \sqrt{1-\left(\frac{\omega}{W_{\rm ebath}}\right)^2},
\end{align}
to avoid the divergence of the bath self-energy in the time representation at $t=t'$.
We note that this bath tends to suppress the SC order, since it reduces the lifetime of quasiparticles.

As for the heat bath attached to the phonon degrees of freedom, we consider a free-boson bath as in the Caldeira-Leggett model,\cite{Caldeira1981}
which is expressed as
\begin{subequations}
\begin{align}
H_{\rm mix,ph}&=\sum_{i,p}\mathcal{V}_{p}(a^\dagger_i+a_{i})(b^\dagger_{i,p}+b_{i,p}),\\
H_{\rm bath,ph}&=\sum_{i,p}\omega_{p}b^\dagger_{i,p}b_{i,p}.
\end{align}
\end{subequations}
Here $b^\dagger_{i,p}$ is the creation operator for a bath phonon with an index $p$, which is attached to the system phonon at site $i$. 
We can trace out the bath phonons in the path-integral formalism, which yields an additional self-energy correction to the phonon Green's function 
in the system,
\begin{align}
\Pi_{i,{\rm bath}}(t,t')&=\sum_{p}\mathcal{V}_{p}^2 D_{0,p}(t,t'),\\
D_{0,p}(t,t')&=-i[\theta_c(t,t')+f_b(\omega_{p})]e^{-i (t-t')\omega_{p}}\nonumber\\
&-i[\theta_c(t',t)+f_b(\omega_{p})]e^{-i(t'-t)\omega_{p}},
\end{align}
where $D_{0,p}$ is the bare bath phonon propagator in equilibrium, $f_b$ is the boson distribution function with the bath temperature $T$,
and $\theta_c(t,t')$ is the Heaviside step function along the Keldysh contour.
We note that the total phonon self-energy includes the contribution from the bath and that from the interaction in $H_{\rm sys}$,
\begin{align}
\Pi_{{\rm tot}}=\Pi_{ {\rm bath}}+\Pi_{{\rm int}}.
\end{align}

Since $\Pi_{{\rm bath}}(t,t')$ is time-translation invariant, we can consider its Fourier components.
The retarded part is 
\begin{align}
\Pi^R_{\rm bath}(\omega)&=\sum_p \mathcal{V}_{p}^2\frac{2\omega_{p}}{(\omega+i0^+)^2-\omega_{p}^2},
\end{align}
and the bath information is contained in its spectrum 
\begin{align}
B_{\rm bath}(\omega)&=- {\rm Im} \Pi_{\rm bath}^R(\omega)\nonumber\\
&=\pi\sum_p \mathcal{V}_{p}^2[\delta(\omega-\omega_{p})-\delta(\omega+\omega_{p})].
\end{align}

When $B_{\rm bath}(\omega)$ is given, the self-energy from the bath can be expressed as
\begin{subequations}
\begin{align}
\Pi^R_{\rm bath}(\omega)&=\frac{1}{\pi}\int\frac{B_{\rm bath}(\omega')}{\omega-\omega'+i0^+}d\omega',\\
\Pi^A_{\rm bath}(\omega)&=\Pi^R_{\rm bath}(\omega)^*,\\
\Pi^K_{\rm bath}(\omega)&=\coth\Bigl(\frac{\beta\omega}{2}\Bigl)[\Pi^R_{\rm bath}(\omega)-\Pi^A_{\rm bath}(\omega)]\nonumber\\
&=-2i \coth\Bigl(\frac{\beta\omega}{2}\Bigl) B_{\rm bath}(\omega).
\end{align}
\end{subequations}

In the following, we consider a bath spectrum of the form 
\begin{align}
B_{\rm bath}(\omega)=\mathcal{V}^2\left\{\frac{\gamma}{(\omega-\omega_{D})^2+\gamma^2}
-\frac{\gamma}{(\omega+\omega_{D})^2+\gamma^2}\right\}.
\end{align}
Here we assume $\gamma=\omega_{D}$, which yields an almost linear structure in $B_{\rm bath}(\omega)$ around $\omega\in[-\omega_{D},\omega_{D}]$.
This can be regarded as an Ohmic bath with a soft cut-off. 
Because of the non-vanishing real part of the bath self-energy, the bath coupling softens the phonon frequency, leading to a stronger attractive interaction.
We put $\gamma_{\rm ph}=\mathcal{V}^2/\omega_D$ in the following.

\subsection{Floquet Green's functions}\label{sec:GF}

We apply
the Keldysh formalism introduced in Sec.~\ref{sec:noneq_form} to a time-periodic NESS
realized under an external periodic field, where we assume that relevant physical observables become periodic in time.
We can take advantage of this to simplify the Keldysh formalism.
For example, the Green's function becomes periodic with respect to the average time $t_{\rm av}$,
\begin{equation}
\begin{split}
G^{\rm R,A,K} (t_{\rm r}; t_{\rm av})=&G^{\rm R,A,K} (t_{\rm r}; t_{\rm av}+\mathscr{T}),\\
D^{\rm R,A,K} (t_{\rm r}; t_{\rm av})=&D^{\rm R,A,K} (t_{\rm r}; t_{\rm av}+\mathscr{T}),
\end{split}
\label{eq:Periodicity}
\end{equation}
where $t_{\rm r}=t-t'$ and $t_{\rm av}=\frac{t+t'}{2}$.
The self-energies have the same property in the NESS.
Hence, the full $t_{\rm av}$-dependence of $G^{\rm R,A,K} (t_{\rm r}; t_{\rm av})$ and $D^{\rm R,A,K} (t_{\rm r}; t_{\rm av})$ contains redundant information. 
We can eliminate this redundancy by considering the Floquet representation of these functions in the following manner.
Let us assume that a function $F(t_{\rm r}; t_{\rm av})$ has the periodicity of Eq.~(\ref{eq:Periodicity}).
The Floquet representation (the Floquet matrix form) of this function is defined as \cite{Aoki2013}
\begin{align}
\label{Fmatrix111}
&{\bf F}_{mn}(\omega)\equiv\\
&\frac{1}{\mathscr{T}}\int^{\mathscr{T}}_0 dt_{\rm av}  \int^\infty_{-\infty} dt_{\rm r} 
e^{i(\omega+\frac{m+n}{2}\Omega) t_{\rm r}}  e^{i(m-n)\Omega t_{\rm av}} \nonumber
F(t_{\rm r};t_{\rm av}).
\end{align}
This representation has some important properties.
At each $\omega$, ${\bf F}_{mn}(\omega)$ can be regarded as a matrix whose indices are  $m$ and $n$ (the Floquet matrix).
In this representation, the convolution of two functions satisfying Eq.~(\ref{eq:Periodicity}) 
in the two-time representation becomes the product of the Floquet matrices at each $\omega$. \cite{Aoki2013}
This simplifies the Dyson equation, Eq.~(\ref{eq:Dyson_time}), to  
\begin{subequations}
\begin{align}
[\underline{\bf G}_0^{-1}(\omega)-\underline{\bf \Sigma}(\omega)]\cdot\underline{\bf G}(\omega)=\underline{\bf I},\label{eq:el_dyson_floq}\\
[\underline{\bf D}_0^{-1}(\omega)-\underline{\bf \Pi}(\omega)]\cdot\underline{\bf D}(\omega)=\underline{\bf I}.\label{eq:ph_dyson_floq}
\end{align}
\end{subequations}
Here $\underline{\bf G}(\omega)$ indicates the Larkin-Ovchinnikov form with each physical component expressed in the Floquet representation, and $\cdot$ is the matrix product with respect to
the Floquet representation and the Larkin-Ovchinnikov form. 
Hence the Dyson equation, which has been originally a Volterra equation in the two-time representation, becomes a matrix inversion problem in the Floquet representation.
In addition, from the definition it follows that $F_{m,n}(\omega)=F_{m+l,n+l}(\omega-l\Omega)$.
We deal with this redundancy by using the reduced Brillouin-zone scheme restricting the frequency to $\omega\in[-\frac{\Omega}{2},\frac{\Omega}{2})$.

When an external periodic field is applied to the electrons, 
we need to incorporate this effect into the bare electron propagator, which acquires a time-periodic structure as in Eq.~(\ref{eq:Periodicity}). 
The corresponding expressions have been discussed in the previous works for the single-band\cite{Tsuji2008} and multi-band\cite{Mikami2016} systems,
so we do not repeat this here.
Instead, we focus on the direct phonon modulation originating from a THz or mid-infrared laser pump, Eq.~(\ref{eq:excitation}).
In this case, we need to first derive the expression for the bare phonon propagator in the presence of such a modulation,
in order to understand the effect of direct phonon excitations and to construct the perturbation theory on top of it.

The free phonon system under the periodic driving is described by the Hamiltonian
\begin{align}
H_{\rm ph}(t)=\omega_{0P}(t) \frac{P^2}{4}+\omega_{0X}(t)\frac{X^2}{4}\label{eq:type2}.
\end{align}
The operator $X$ in the Heisenberg representation satisfies the equation
\begin{align}
\left[-\partial_t^2+\frac{\partial_t \omega_{0P}(t)}{\omega_{0P}(t)}\partial_t-\omega_{0P}(t)\omega_{0X}(t)\right]X(t)=0.
\end{align}
From this, we find that  
\begin{align}
D^{-1}_0(t,t')\equiv\frac{-\partial_t^2+\frac{\partial_t \omega_{0P}(t)}{\omega_{0P}(t)}\partial_t-\omega_{0P}(t)\omega_{0X}(t)}{2\omega_{0P}(t)}\delta(t-t') \label{eq:D0_trep}
\end{align}
satisfies
\begin{subequations}\label{eq:D0_eq}
\begin{align}
\int d\bar{t} D^{-1}_0(t,\bar{t})D_0^R(\bar{t},t')&=\delta(t-t'),\\
\int d\bar{t} D^{-1}_0(t,\bar{t})D_0^K(\bar{t},t')&=0,\\
\int d\bar{t} D^{-1}_0(t,\bar{t})D_0^A(\bar{t},t')&=\delta(t-t').
\end{align}
\end{subequations}
Hence, in the Dyson equation for the full phonon Green's function, Eq.~(\ref{eq:ph_dyson_time}), we can use $\underline{D}_0^{-1}(t,t')=\underline{I}D^{-1}_0(t,t')$.
In the Floquet representation, $D^{-1}_0(t,t')$ becomes 
\begin{align}
[{\bf D}_0]^{-1}_{mn}(\omega)=\frac{(\omega+n\Omega)(\omega+m\Omega)[{\boldsymbol \omega}_{0P}^{\rm inv}]_{mn} -[{\boldsymbol \omega}_{0X}]_{mn}}{2}, \label{eq:D0_floq}
\end{align}
where
\begin{subequations}
\begin{align}
[{\boldsymbol \omega}_{0X}]_{mn}&=\frac{1}{\mathscr{T}}\int^{\mathscr{T}}_{0} dt_{\rm av} e^{i(m-n)\Omega t_{\rm av}}\omega_{0X}(t_{\rm av}),\\
[{\boldsymbol \omega}_{0P}^{\rm inv}]_{mn}&=\frac{1}{\mathscr{T}}\int^{\mathscr{T}}_{0} dt_{\rm av} \frac{e^{i(m-n)\Omega t_{\rm av}}}{\omega_{0P}(t_{\rm av})}.
\end{align}
\end{subequations}
We note that the inverse of the Floquet matrix of the retarded part, $[{\bf D}^R_0(\omega)]^{-1}$, is equal to $[{\bf D}_0(\omega+i0^+)]^{-1}$.
In the numerical implementation, we solve the Dyson equation for the phonon Green's function using Eq.~(\ref{eq:D0_floq}) and Eq.~(\ref{eq:ph_dyson_floq}).

Now we focus on more specific situations described by the driving protocols of type 1 [Eq.~\eqref{eq:type1}] and type 2 [Eq.~\eqref{eq:Komnik}]. 
For the excitation of type 1 [Eq.~(\ref{eq:type1})], the inverse of the bare phonon Green's function is obtained by substituting 
\begin{subequations}
\begin{align}
[{\boldsymbol \omega}_{0X}]_{mn}&=\omega_0\delta_{m,n}+\frac{\Delta \omega_0}{2}(\delta_{m,n+1}+\delta_{m,n-1}),\\
[{\boldsymbol \omega}^{\rm inv}_{0P}]_{mn}&=\frac{1}{\omega_0}\delta_{m,n},
\end{align}
\end{subequations}
into Eq.~(\ref{eq:D0_floq}).
Since the equation of $D^{R}_0(t,t')$ turns out to be the Mathieu's differential equation (see also Sec.~\ref{sec:para_ins}), we can express it as a linear combination of 
its two Floquet solutions.

For the excitation of type 2 [Eq.~(\ref{eq:Komnik})], the inverse of the bare phonon Green's function is obtained from  
\begin{align}
[{\boldsymbol \omega}_{0X}]_{mn}&=\omega_0\delta_{m,n}+\frac{\Delta \omega_0}{2}(\delta_{m,n+1}+\delta_{m,n-1}),\\
[{\boldsymbol \omega}^{\rm inv}_{0P}]_{mn}&=\frac{1}{\omega_0}\sum_{l=0}^\infty\frac{(2l+r)!}{(l+r)!l!} \left(\frac{-\Delta \omega_0}{2\omega_0}\right)^{2l+r}\\
&=\frac{1}{\omega_0} \left(\frac{-\Delta \omega_0}{2\omega_0}\right)^{r} {}_2 F_1 \left(\frac{1+r}{2},\frac{2+r}{2};1+r;\frac{\Delta\omega_0^2}{\omega_0^2}\right),\nonumber
\end{align}
where we put $r=m-n$ and ${}_2F_1(\alpha,\beta;\gamma;z)$ is the hypergeometric function.
One can also easily express $D^{R}_0$ in the time domain as 
\begin{align}
D_0^R(t,t')=\theta(t,t')i[e^{i(S(t)-S(t'))}-e^{-i(S(t)-S(t'))}],
\end{align}
where $S(t)\equiv t\omega_0+\frac{\Delta \omega_0}{\Omega}\sin(\Omega t)$.

\subsection{Dynamical mean-field theory}\label{sec:FDMFT}

In order to evaluate the full Green's functions, we employ the dynamical mean-field theory (DMFT), \cite{Metzner1989,Georges1992,Georges1996,Aoki2013}
which becomes exact in the limit of infinite spatial dimensions. \cite{Georges1996}
The idea of DMFT is to map the lattice problem to an effective impurity problem.
In the present case of the Holstein model with thermal baths, the form of the effective impurity model in the path-integral formalism is 
\begin{align}
S_{\rm imp}&=i\int_{{\mathcal C}_K} dt dt' \Psi^{\dagger}(t) (\hat{\mathcal{G}}^{-1}_{0} (t,t')-\hat{\Sigma}_{\rm bath}(t,t')) \Psi(t')\nonumber\\
&+i\int_{{\mathcal C}_K} dt dt' X(t) \frac{D_0^{-1}(t,t')-\Pi_{\rm bath}(t,t')}{2}  X(t')\nonumber\\
&-ig\int_{{\mathcal C}_K}dt X(t)\Psi^{\dagger}(t) \hat{\sigma}_3 \Psi(t),
\end{align}
where 
\begin{align}
\hat{\mathcal{G}}^{-1}_{0} (t,t') =[i\partial_{t}\hat{I}+\mu\hat{\sigma}_3]\delta_{{\mathcal C}_K}(t,t')-\hat{\Delta}(t,t'),
\end{align}
$\int_{\mathcal{C}_K}$ denotes an integral on the Keldysh contour, $\Psi^\dagger(t)\equiv[c_{\uparrow}^\dagger(t),c_{\downarrow}(t)]$ a Nambu spinor, and a hat symbol a $2\times2$ matrix in the Nambu form.
In this expression,  the thermal baths and $P$ (phonon momentum) in the impurity problem have been integrated out.\cite{Murakami2016}
The hybridization function $\hat{\Delta}$ is determined such that the impurity Green's function, $\hat{G}_\text{imp}(t,t')=-i\langle T_{\mathcal{C}_K} \Psi(t) \Psi^\dagger(t') \rangle$, and the impurity self-energy $\hat{\Sigma}_{\rm imp, int}$ are identical to the local lattice Green's function for the electrons, $\hat{G}_{\rm loc}(t,t')=-i\langle T_{\mathcal{C}_K} \Psi_i(t) \Psi_i^\dagger(t') \rangle$,  and the local lattice self-energy, respectively.

Since we are considering a time-periodic NESS, the effective impurity model also has the same time periodicity, where the hybridization function, $\hat{\Delta}$,
is assumed to show the periodic behavior of Eq.~(\ref{eq:Periodicity}). 
Hence the Dyson equation involved in the solution of the effective impurity model can be dealt with as an inverse problem of the Floquet matrix.
Precisely speaking, the matrix to invert involves three indices, i.e., those of the Floquet, Larkin-Ovchinnikov, and Nambu forms. 

\subsubsection{Migdal approximation}
In order to solve the effective impurity model, we employ the self-consistent Migdal approximation,
\cite{Murakami2015,Murakami2016,Murakami2016b,Bauer2011,Freericks1994,Capone2003,Hague2008,Leeuwen2015,Pavlyukh2016} 
which is justified when the phonon frequency $\omega_0$ is small enough compared to the electron bandwidth.\cite{Bauer2011,Freericks1994,Capone2003,Hague2008}
Here, the electron self-energy ($\hat{\Sigma}$) 
and phonon self-energy ($\Pi$) are given by
\begin{equation}
\begin{split}
\hat{\Sigma}_{\rm int}(t,t')&=ig^2D_{\rm imp}(t,t')\hat{\sigma}_3\hat{G}_{\rm imp}(t,t')\hat{\sigma}_3,\\
\Pi_{\rm int}(t,t')&=-ig^2{\rm Tr}[\hat{\sigma}_3 \hat{G}_{\rm imp}(t,t')\hat{\sigma}_3\hat{G}_{\rm imp}(t',t)],
\label{eq:Migdal}
\end{split}
\end{equation}
where we do not explicitly write the Hartree term.
Remember that in DMFT $\hat{\Sigma}_{\rm int}(t,t')$ is identified with the lattice self-energy of the electrons at each momentum.
The Hartree term is proportional to $\langle X(t)\rangle$, and, in the half-filled case, which we focus on in this paper, it is exactly zero.
Away from half-filling, $\langle X(t)\rangle$ can be finite and time dependent. 
Still, since we are considering homogeneous excitations, this term can be regarded as a time-dependent chemical potential and does not affect the dynamics of relevant physical quantities without couplings to the B\"{u}ttiker-type bath.
The self-consistent Migdal approximation is a conserving approximation, which allows one to trace the energy flow from the system to the thermal baths.

\subsubsection{Observables}

We summarize the physical observables that are used to discuss the results in the paper.
Let us denote the Heisenberg representation as $\mathcal{O}(t)=\mathcal{U}_\text{tot}(t_0,t)\mathcal{O}\mathcal{U}_\text{tot}(t,t_0)$, where $\mathcal U_{\rm tot}(t,t_0)$ is the unitary evolution operator from time $t_0$ to $t$.

The change of the system energy can be expressed as
\begin{align}
\frac{d E_{\rm sys}(t)}{dt}=\langle \partial_t H_{\rm sys}(t)\rangle +i\langle [H_{\rm mix}(t),H_{\rm sys}(t)]\rangle.\label{eq:eq_esys}
\end{align}
The first term represents the work done by the external field, which we express as $W(t)$.
The second term represents the energy dissipation from the system to the bath,
which we express as $-I(t)$.
If the system is in a time-periodic NESS, the average energy injected from the field and dissipated to the baths should be the same,
\begin{align}
\overline{ W}=\bar{I},
\end{align}
where the overline indicates the time average over one period, $\frac{1}{\mathscr{T}}\int_0^{\mathscr{T}} dt $.
We also note that the energy dissipated from the system should be equal to the energy injected to the bath.
This leads to 
\begin{align}
\bar{I}= i\overline{\langle [H_{\rm mix}(t),H_{\rm bath}(t)]\rangle}, \label{eq:Hrada_sasa}
\end{align}
see also Appendix~\ref{sec:Harada} for more details.

In the following, we give the explicit expression for each term in Eq.~(\ref{eq:eq_esys}) using the self-energies and Green's functions.\\

(1) The work done by the field is 
\begin{align}
&W(t)= 
\begin{cases}
-\frac{\Delta \omega_0\Omega \sin(\Omega t)}{4}\sum_i \langle X_i^2(t) \rangle & \text{(Type 1)},\\
-\frac{\Delta \omega_0\Omega \sin(\Omega t)}{4}\sum_i (\langle P_i^2(t) \rangle +\langle X_i^2(t)) \rangle & \text{(Type 2)},\\
\end{cases}
\end{align}
where $\langle X_i^2(t) \rangle$ can be obtained from the lesser part of the phonon Green's function. 

(2) The dissipated energy is
\begin{align}
I(t)&=I_{\rm ph}(t)+I_{\text {el}}(t)\nonumber\\
&=-i\langle [H_{\rm ph\_mix}(t),H_{\rm sys}(t)]\rangle-i\langle [H_{\rm el\_mix}(t),H_{\rm sys}(t)]\rangle,
\end{align}
where 
\begin{align}
I_{\rm ph}(t)=i\sum_i \partial_t[D_i*\Pi_{i,\rm ph\_bath}]^<(t,t')|_{t'=t}
\end{align}
and
\begin{align}
I_{\text {el}}(t)&=\sum_{i}{\rm Tr}[\hat{\Sigma}_{\rm el\_bath}\circledast\hat{\Delta}_{i}\circledast \hat{G}_{ii}]^<(t,t)+H.c.\nonumber\\
&-\mu\sum_{i}{\rm Tr} [\sigz\cdot\hat{\Sigma}_{\rm el\_bath}\circledast \hat{G}_{ii}]^<(t,t)+H.c\nonumber\\
&-\sum_{i}{\rm Tr}[\hat{\Sigma}_{{\rm int},i} \circledast\hat{G}_{ii} \circledast\hat{\Sigma}_{\rm el\_bath}]^<(t,t)+H.c.
\end{align}
Here the symbol $*$ indicates the convolution along the Keldysh contour $\mathcal C_K$ and the symbol $\circledast$ represents a multiplication of the 2$\times$2 Nambu matrices and the convolution i.e.,
$[\hat{A}\circledast\hat{B}]_{\alpha,\alpha'}(t,t')=\sum_\gamma\int_{\mathcal C}d\bar{t} A_{\alpha,\gamma}(t,\bar{t})B_{\gamma,\alpha'}(\bar{t},t')$.
${\rm Tr}$ refers to the trace of Nambu matrices.\\

The system observables (that $H_{\rm sys}$ involves) can be expressed in the following manner, in analogy to the Kadanoff-Baym formalism.\cite{Murakami2015}

\underline{1. Kinetic energy}
\begin{align}
E_{\rm kin}(t)&=\sum_{i,j,\sigma}-v_{i,j}\langle c_{i,\sigma}^{\dagger}(t)c_{j,\sigma}(t)\rangle\nonumber\\
&=-i{\rm Tr}[\hat{\Delta}\circledast \hat{G}_{\rm loc}]^<(t,t).
\end{align}

\underline{2. Interaction energy}
\begin{align}
E_{\rm nX}(t)&=g\sum_{i}\langle X_i(t)\hat{\Psi}^\dagger_i\hat{\sigma}_3\hat{\Psi}_{i}(t)\rangle=-i{\rm Tr}[\hat{\Sigma}_{\rm int}\circledast \hat{G}_{\rm loc}]^<(t,t).
\end{align}

\underline{3. Phonon energy}
\begin{align}
E_{\rm ph}(t)=\frac{\omega_{0X}(t)}{4}\sum_i (iD_i^<(t,t))+\frac{\omega_{0P}(t)}{4}\sum_i (iD_{\rm PP,i}^<(t,t)).
\end{align}
Here $D_{\rm PP}(t,t')=-i\langle T_c P(t)P(t')\rangle$.
This can be obtained by taking derivatives of $D(t,t')$.\cite{Murakami2015}
The Larkin-Ovchinnikov form of this function becomes
\begin{align}
\underline{D_{PP}}(t,t')&=\frac{1}{\omega_{0P}(t)\omega_{0P}(t')}\partial_{t}\partial_{t'} \underline{D}(t,t')-\frac{2\delta(t-t')}{\omega_{0P}(t)}\underline{I}. 
\end{align}
We note that the derivative in the two-time representation becomes a multiplication of factors depending on $\omega,m$ and $n$ in the Floquet representation
, which can be used for the numerical implementation.

Another important quantity is the optical conductivity, i.e.
 $\sigma_{\alpha\alpha}(t,t')\equiv \delta \langle j_{\alpha} (t)\rangle/\delta E_\alpha(t')$, where $\langle j_{\alpha} (t)\rangle$ is the current for the direction $\alpha$, $E_\alpha(t')$ is the probe field, and
 $\delta$ indicates the functional derivative.\cite{Aoki2013}
This can be expressed in terms of the susceptibility $\chi_{\alpha\alpha}(t,t')\equiv \delta \langle j_{\alpha} (t)\rangle/\delta A_\alpha(t')$ as 
\begin{align}
\sigma_{\alpha\alpha}(t,t')&=-\int^t_{t'}\chi_{\alpha\alpha}(t,\bar{t})d\bar{t}.
\end{align}
$\chi_{\alpha\alpha}(t,t')$ consists of the diamagnetic term $\chi^{\rm dia}_{\alpha\alpha}(t,t')$ and the paramagnetic term $\chi^{\rm para}_{\alpha\alpha}(t,t')$.\cite{Aoki2013}
In principle, the paramagnetic term includes the vertex correction, but in the present case, since the phonon driving  does not break the parity symmetry (${ \bf k}\leftrightarrow {\bf -k}$), its contribution vanishes. In the end, we obtain the following expression in the Nambu form,
\begin{subequations}
\begin{align}
\chi^{\rm dia}(t,t')&=i\delta(t-t') \int d\epsilon \frac{d\Phi(\epsilon)}{d\epsilon}  {\rm Tr}[\hat{\sigma}_3\hat{G}_{\epsilon}^<(t,t)],\\
 \chi^{\rm para}(t,t')&=i \int d\epsilon \Phi(\epsilon)
     \Bigl\{{\rm Tr}[\hat{G}^R_{\epsilon}(t,t')\hat{G}^<_{\epsilon}(t',t)\nonumber\\
     &\;\;\;\;\;\;\;\;\;\;\;\;\;\;\;\;\;\;\;\;\;\;+{\rm Tr}[\hat{G}^<_{\epsilon}(t,t')\hat{G}^A_{\epsilon}(t',t)]\Bigl\},
\end{align}
\end{subequations}
where $\epsilon$ denotes the energy of noninteracting electrons, $\Phi(\epsilon)=\frac{1}{\Omega_{\rm vol}}\sum_{{\bf k},\sigma} \left( \frac{\partial\epsilon_{\bf k}}{\partial k_\alpha}\right)^2\delta(\epsilon-\epsilon_{\bf k})$, $\frac{d\Phi(\epsilon)}{d\epsilon}=\frac{1}{\Omega_{\rm vol}}\sum_{\bf k}\frac{\partial^2\epsilon_{\bf k}}{\partial k_\alpha^2} \delta(\epsilon-\epsilon_{\bf k})$, and $\Omega_{\rm vol}$ is the volume of the system. 
 In particular, for the Bethe lattice, $\Phi(\epsilon)=(N/3d)[(W/2)^2-\epsilon^2]\rho_0(\epsilon)$ is used, where $N$ is the system size and 2$d$ the coordination number.\cite{Bethe_trans}

In the NESS,  $\sigma(t,t')$ and $\chi(t,t')$ become time periodic, and their  Floquet expressions are connected through 
\begin{align}
\boldsymbol{\sigma}_{mn}(\omega)=\frac{\boldsymbol{\chi}_{mn}(\omega)}{i(\omega+n\Omega)}.
\end{align}
To obtain an intuitive picture, let us imagine that $E_\alpha(t)=\delta E_0e^{-i\nu t}$ is applied as a probe field.
The induced current is 
\begin{align}
\delta j_\alpha(t)&=\delta E_0\sum_{m} e^{-i(\nu+m\Omega)t} \boldsymbol{\sigma}_{m0}(\nu).\label{eq:cond}
\end{align}
We note that in this expression we unfold the restricted zone scheme for $\boldsymbol{\sigma}^{ij}_{m0}(\nu)$ by using $\boldsymbol{\sigma}_{m0}(\nu)=\boldsymbol{\sigma}_{m+l,l}(\nu-l\Omega)$. Equation~\eqref{eq:cond} means that the response includes $e^{-i (\nu+m\Omega) t}$ components in addition  to $e^{-i\nu t}$. Previous studies \cite{Tsuji2009,Mikami2016}  have mainly focused on the diagonal component ($m=0$), while in the present work, we will also discuss the off-diagonal components.

\subsection{Breakdown of the Floquet Green's function method}\label{sec:para_ins}
It is important to remark that there are instabilities in the system under the excitation Eq.~(\ref{eq:type1}).
Without the heat bath and the el-ph coupling, the equation for the phonon Green's function [Eq.~(\ref{eq:D0_eq}) combined with Eq.~(\ref{eq:type1})]
becomes the so-called Mathieu's differential equation. 
It is known that the solution of this equation is unstable around $\frac{n}{2}\Omega=\omega_0$, with $n$ a natural number \cite{}.
In this unstable regime, the amplitude of oscillations in the solution of the equation monotonically increases, and 
there is no well-defined Floquet form of the Green's function since we cannot define the Fourier components. 

 \begin{figure}[t]
  \centering
      \vspace{0.cm}
   \includegraphics{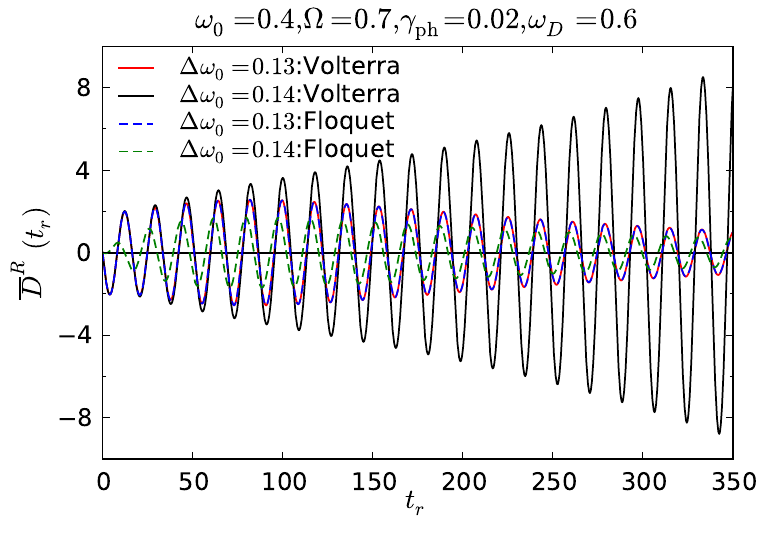}
  \caption{Comparison between solutions of the Dyson equation obtained from the Volterra equation, Eq.~(\ref{eq:ph_dyson_time}), and the Floquet representation, Eq.~(\ref{eq:ph_dyson_floq}).
 $\bar{D}^R(t_{\rm r})$ is $D^R(t_{\rm r};t_{\rm av})$ averaged over $t_{\rm av}$. Only the self-energy correction from the phonon bath is considered.}
  \label{fig:parametric_instability}
\end{figure}

The heat bath and the el-ph coupling introduce a damping effect through the self-energy corrections. 
Even though the damping effect can reduce the unstable regime,\cite{Zerbe1994,Zerbe1995} 
the Green's function still exhibits an instability when the strength of the external field is large enough.
In such cases the Floquet Green's function cannot be defined and the Floquet formalism becomes inapplicable.
Physically, this means that the NESS, if any, is unstable against infinitesimally small perturbations to the phonon part.

In Fig.~\ref{fig:parametric_instability}, we demonstrate that even with the phonon bath the behavior of the retarded part of the phonon Green's function
changes qualitatively as a function of driving amplitude. 
The correct results are obtained by solving the Dyson equation in the time domain (differential-integral Volterra equation), Eq.~(\ref{eq:ph_dyson_time}),
which produces either a stable or unstable solution depending on the excitation parameter.
For comparison, we solve the Dyson equation in the Floquet form, Eq.~(\ref{eq:ph_dyson_floq}), and carry out an inverse transformation to the two-time representation. 
These two ways of solving the Dyson equation give the same solution, when the solution is stable.
However, we have found an unphysical solution for the unstable regime in the Floquet form that is different from the transient solution of the Volterra equation in the time domain.
In particular, the unphysical solution violates the initial condition (the sum rule in the $\omega$ space) $\partial_t D_0^R(t,t')|_{t'=t}=-2\omega_0$, see also Sec.~\ref{sec:D0}.

Therefore, when using the Floquet form of the Green's function, one has to be careful about the unstable regime, where the system 
cannot reach a physically stable NESS. The excitation Eq.~(\ref{eq:type1}) at the parametric resonance $\frac{\Omega}{2}=\omega_0$ is potentially in such a regime.
However, there is no such problem in the calculation of the transient time evolution on the Kadanoff-Baym contour. 

\section{Results}\label{sec:results}
 \begin{figure}[t]
  \centering
      \vspace{0.cm}
   \includegraphics[width=70mm]{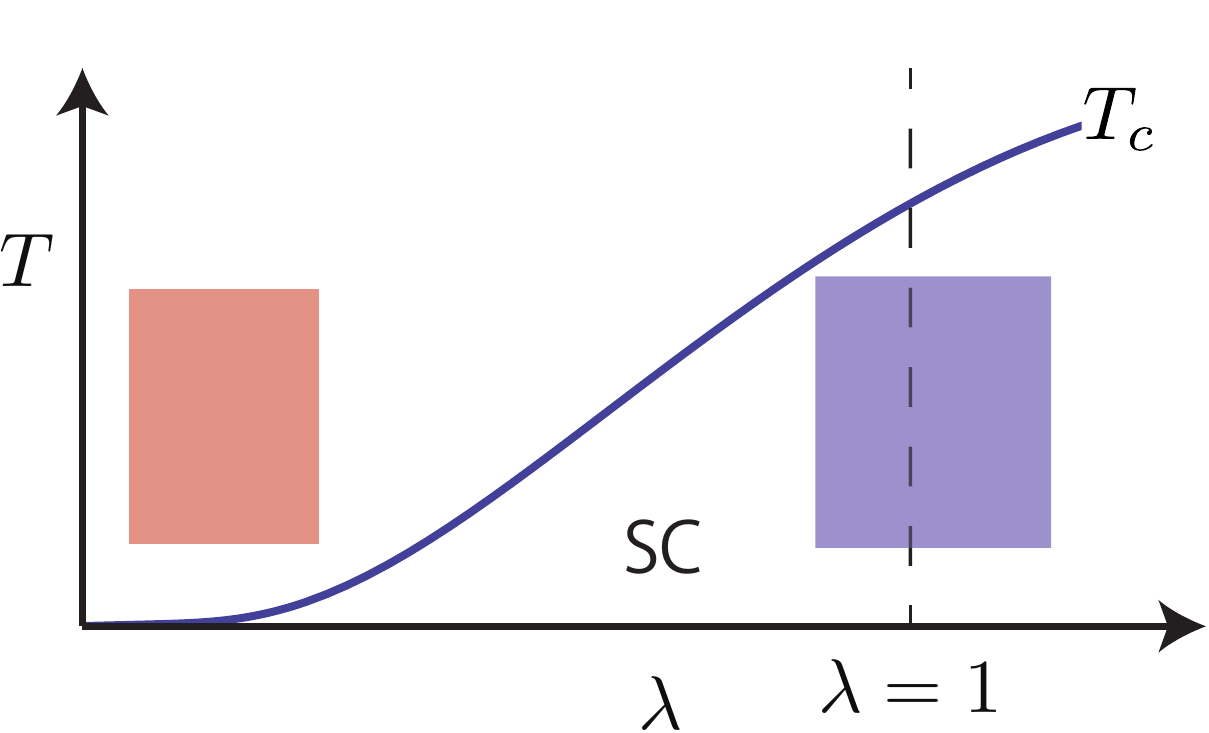}
  \caption{Schematic picture of the parameter regimes discussed in this paper. The dashed line represents $\lambda=1$.}
  \label{fig:regime}
\end{figure}
In this paper, we investigate the effect of parametric phonon driving on superconductivity (SC) both in the weak coupling ($\lambda \ll 1$, the red region in Fig.~\ref{fig:regime}) and strong coupling  ($\lambda\sim1$, the blue region in Fig.~\ref{fig:regime}) regimes. 
 In this paper, we use the semi-elliptic density of states, $\rho_0(\omega)=\frac{1}{2\pi v^2_{\ast}} \sqrt{4v^2_\ast-\epsilon^2}$, i.e. the Bethe lattice.
We set $v_\ast=1$, i.e. the electron bandwidth $W$ is 4, and we focus on the half-filled case.
We choose the bare phonon frequency $\omega_0=0.4$, which is much smaller than the electron bandwidth.
If we take the bandwidth as $0.4$ eV, the bare phonon frequency is $40$ meV, corresponding to a period of about $100$ fs.
The gap size of the SC spectrum around $\lambda\sim1$ is approximately $10$ meV, i.e. this period is about $400$ fs.
Our unit temperature corresponds to $0.1$ eV or 1160 K.

\subsection{Effective attractive interaction in the weak coupling limit : Bare phonon propagator}\label{sec:D0}
 \begin{figure*}[t]
  \centering
      \vspace{0.cm}
   \includegraphics[width=160mm]{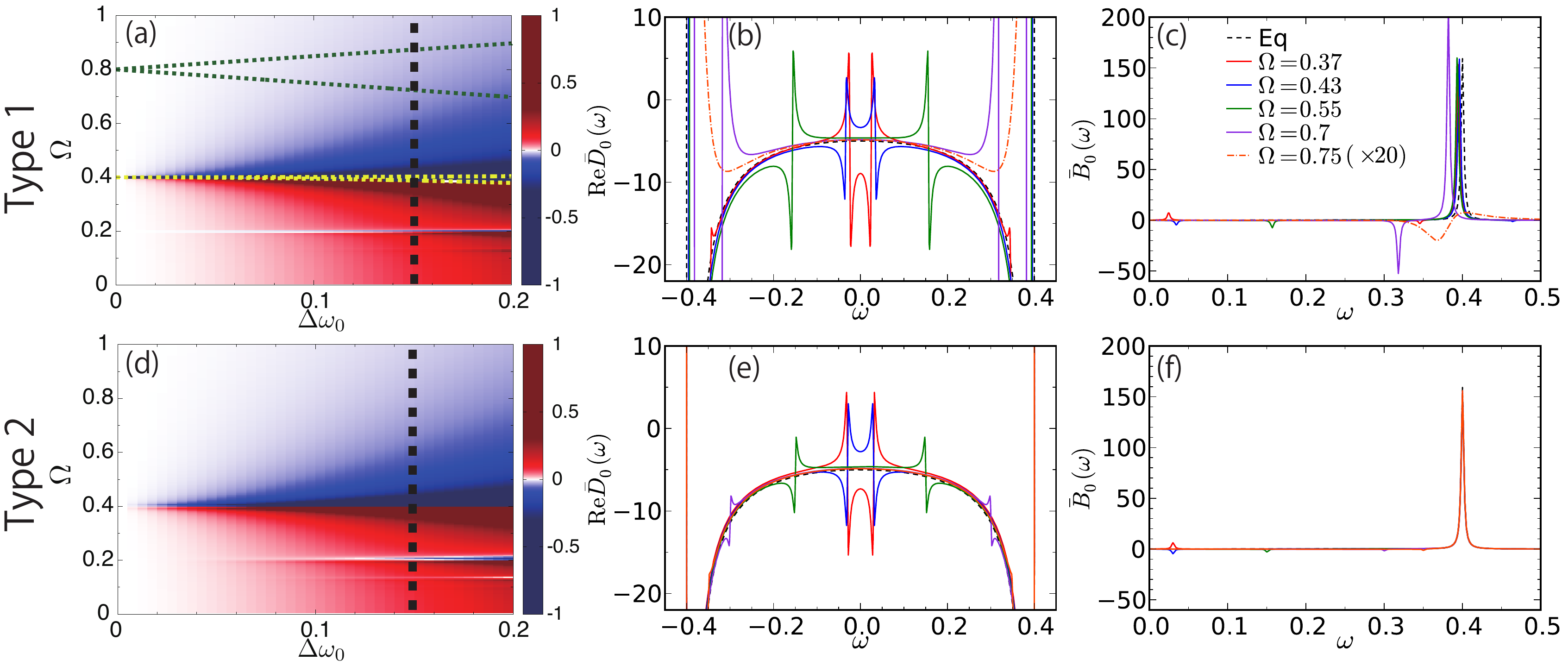}
  \caption{Properties of $\bar{D}^R_0(\omega)$ for the type 1 excitation (a)(b)(c) and the type 2 excitation (d)(e)(f).
  Here we choose $\omega_0=0.4$. (a)(d) show the change of the effective attractive interaction expressed as $({\rm Re} \bar{D}^R_0(0)-{\rm Re }D_{\rm 0,eq}(0))/{\rm Re } D_{\rm 0,eq}(0)$. The yellow and green dotted lines in (a) indicate the boundaries of the unstable regimes of the Mathieu equation around $\Omega=2\omega_0$ and $\Omega=\omega_0$. (b)(c)(e)(f) show the excitation-frequency dependence of $\bar{D}^R_0(\omega)$ along the black dashed line in (a)(d). (b)(e) show  ${\rm Re} \bar{D}^R_0(\omega)$  and (c)(f) show $\bar{B}_0(\omega)\equiv-\frac{1}{\pi}{\rm Im} \bar{D}^R_0(\omega)$. The dotted color lines ($\Omega=0.75$) in (b)(c) correspond to an unphyscial solution.
  The spectra are broadened by adding a damping term $-\gamma \partial_t\delta(t-t')$ with $\gamma=0.005$ in Eq.~(\ref{eq:D0_trep}).
  }
  \label{fig:D0_info}
\end{figure*}

In this section, we discuss how the effective attractive interaction mediated by phonons, which is responsible for the SC,
 is modulated under the phonon driving in the weak coupling limit. 
 Our discussion is based on the Migdal-Eliashberg theory, where the retarded attractive interaction mediated by the phonons is represented by the phonon propagator, $g^2D^R(\omega)$, 
 and  $g^2D^R(0)$ is conventionally taken as a measure for the strength of the attractive interaction.
 In the weak coupling limit, the renormalization of the phonon (the self-energy correction) is small and $g^2D^R_0(\omega)$ serves as a measure of the effective attractive interaction.
 With this idea in mind, the time-averaged $\bar{D}^R_0(0)=\frac{1}{\mathscr{T}}\int_0^\mathscr{T} dt_{\rm av} \int^\infty_{-\infty} dt_{\rm r} D^R_0(t_{\rm r};t_{\rm av})$ has been analyzed for the type 2 excitation in Ref.~\onlinecite{Komnik2016}
 
The effect of the type 1 excitation, on the other hand, has been discussed in Ref.~\onlinecite{Knap2016} using the polaron picture and the Lang-Firsov transformation.
The polaron approximation works well when the phonon timescale is comparable or faster than the electronic timescale. \cite{Murakami2013a}
Within this framework, the hopping of polarons is renormalized from the bare electron hopping parameter by the Frank-Condon factor, $\exp(-g^2/\omega_0^2)$, and the effective attractive interaction is $-\lambda_0=\frac{2g^2}{\omega_0}$. 
However, in many materials the phonon frequencies are much smaller than the bandwidth, in which case the Migdal-Eliashberg theory is applicable.
We note that these two different theories can lead to different predictions concerning the behavior of the SC order. 
The isotope effect serves as an example.
By the substitution of isotopes, usually $\omega_0$ decreases as $\propto\frac{1}{\sqrt {M}}$ ($M$ is the mass of the atom), while $\lambda_0$ does not change. 
 In the polaron picture, the system moves to the stronger coupling regime, since the Frank-Condon factor decreases.
 The convex shape of the superconducting phase boundary in the weak-coupling regime (see Fig.~\ref{fig:regime}) thus implies an increase of the transition temperature.
 On the other hand, the BCS theory, to which the Migdal-Eliashberg theory reduces in the weak coupling regime, predicts the opposite behavior. 
 Here, the transition temperature is proportional to $\omega_0\exp(-\frac{1}{\lambda_0 \rho_0(0)})$, which decreases 
 if $\omega_0$ is reduced by the substitution of isotopes.
Hence for realistic phonon frequencies we should discuss the effects of parametric phonon excitations within the Migdal-Eliashberg theory.

 In Fig.~\ref{fig:D0_info}, we show the numerical results for the time-averaged phonon propagator, $\bar{D}^R_0(\omega)=\frac{1}{\mathscr{T}}\int_0^\mathscr{T} dt_{\rm av} \int^\infty_{-\infty} dt_{\rm r} D^R_0(t_{\rm r};t_{\rm av})e^{i\omega t_{\rm r}}$, which corresponds to the diagonal components in the Floquet representation, for both types of excitations. 
In the evaluation, we formally  invert the Floquet matrix of $D^{\rm R-1}_0$.
Therefore, the results also include unphysical ones as we explained in Sec.~\ref{sec:para_ins}, but we can detect the unstable regime from the Mathieu's equation.
 In Fig.~\ref{fig:D0_info}(a)(d), we plot the change in the effective attractive interaction expressed as $({\rm Re} \bar{D}^R_0(0)-{\rm Re }D_{\rm 0,eq}(0))/{\rm Re } D_{\rm 0,eq}(0)$ in the plane of $\Delta \omega_0$ and $\Omega$.
 It turns out that both excitations result in a similar behavior. 
 There is a strong enhancement of the attractive interaction $\bar{D}^R_0(0)$ at $\Omega \lesssim \omega_0$.
On the other hand, for $\Omega \gtrsim \omega_0$, the attractive interaction is strongly suppressed and it can even become repulsive as is analytically shown in Ref.~\onlinecite{Komnik2016}.

This behavior is related to the position of the Floquet sidebands of the phonons in the phonon spectrum, $\bar{B}_0(\omega)\equiv-\frac{1}{\pi}{\rm Im} \bar{D}^R_0(\omega)$.
At $\Omega \lesssim \omega_0$, the first replica peak ($\sim \omega_0-\Omega$) is just above $\omega=0$. 
At $\Omega \gtrsim \omega_0$, the first side peak moves to $\omega<0$ and there emerges a negative peak in the spectrum at $\omega>0$,
which is the first sideband of the phonon peak at $-\omega_0$, in the time-averaged spectrum $\bar{B}_0(\omega)$, see Fig.~\ref{fig:D0_info} (c)(f).
Since $\bar{D}^R_0(\omega)=\int d\omega'\frac{\bar{B}_0(\omega')}{\omega-\omega'+i0^+}$, in the former case ($\Omega \lesssim \omega_0$) the sideband acts as an additional phonon coupled to electrons with a positive frequency, which leads to an additional attractive interaction. In the latter case, the sideband resembles a negative frequency phonon attached to electrons, which yields an effective repulsive interaction.
There are analogous singularities around $\Omega=\omega/n$ related to the zero crossing of the $n$-th Floquet sideband of the phonons.
Since the spectral weight of such high-order sidebands is small, a substantial effect only appears when the excitation strength $\Delta \omega_0$ is large enough.
Even though the strong enhancement of the attractive interaction is restricted to some energy window smaller than $\omega_0$, see Fig.~\ref{fig:D0_info}(b)(e), 
in the weak coupling limit, the effect of the energy window is smaller than that of the strength of the attractive interaction 
as can be seen from the BCS expression for the transition temperature. The former is a prefactor, while the latter is in the exponential.
Hence it is natural to expect that this may lead to an enhancement of SC assuming that the electron distribution is the same as in equilibrium.
 
Here we note that the unstable regime in the case of the type 1 excitation is located around $\Omega=\frac{2\omega_0}{m}$.
The boundary of the unstable region is depicted by colored dotted lines in Fig.~\ref{fig:D0_info}(a).
Therefore, when $m=2n$, the unstable regime overlaps with the $\Omega$-regime in which $\bar{D}^R_0(0)$ exhibits drastic changes.
 On the other hand, when $m=2n-1$ the existence of the unstable regime 
 does not leave any trace in 
 $\bar{D}^R_0(0)$.
It turns out, instead, that the effect of the instability emerges in the finite frequency part of $\bar{D}^R_0(\omega)$.
As can be seen in the result for $\Omega=0.7$ and $\Omega=0.75$ in Fig.~\ref{fig:D0_info} (b)(c), there is a large change in $\bar{D}^R_0(\omega)$ around $\omega\sim \omega_0$. 
The results for $\Omega=0.75$ are unphysical as discussed in Sec.~\ref{sec:para_ins}, and the spectrum shows dominant negative weight at $\omega>0$, which leads to the 
violation of the sum rule, $2\omega_0=\int^\infty_{-\infty}\omega \bar{B}(\omega)d\omega$. 
Based on the behavior of
 $\bar{D}^R_0(0)$, we thus do not expect any enhancement of SC around the parametric resonant regime at $m=2n-1$ for the type 1 excitation, and 
the analysis of $\bar{D}^R_0(0)$ results in a picture which differs from the enhancement of SC at the parametric resonance discussed in Ref.~\onlinecite{Knap2016}.
For the type 2 excitation, there is no unstable solution, and we do not observe any large change in $\bar{D}^R_0(\omega)$ near $\omega=\omega_0$, see Fig.~\ref{fig:D0_info} (e)(f).

Even though the increased attractive interaction $g^2\bar{D}^R_0(0)$ seems a plausible scenario for the enhancement of SC,  
this naive expectation ignores the following issues: 
 (i) The system gains energy from the external field, which results in heating and nonthermal distribution functions.  
 (ii) When the electron-phonon coupling is not small, the phonon spectrum is renormalized and broadened, and the behavior of $D$ might be different from $D_0$.
 (iii) $\bar{D}^R_0(0)$ is an average of $D^R_0(t_{\rm r};t_{\rm av})$, and hence includes information of infinite past and infinite future times.
Therefore, it is not clear to what extent $\bar{D}^R_0(0)$ is relevant to the transient dynamics of the superconducting order.

In the following sections, using the nonequilibrium DMFT and the Holstein model, 
we provide a comprehensive study of the effects of the parametric phonon driving on SC which takes the above issues into account. 
We also note that it turns out that the type 1 and type 2 excitations give the same results on a qualitative level. 
Therefore, in the main text, we only show the results for the type 1 excitation, and present the results for the type 2 excitation in Appendix \ref{appendix:type2}.
\subsection{Nonequilibrium steady states under phonon driving}\label{sec:NESS}
In this section, we study the effects of the phonon driving on SC in the strong electron-phonon coupling regime by studying the nonequilibrium steady states.
In the following we use, as a representative set of parameters for the strong coupling regime, $g=0.41,\omega_0=0.4,\beta=120$ and the bath parameters $\gamma_{\rm ph}=0.02,\omega_D=0.6,\gamma_{\text {el}}=0.005,W_{\rm ebath}=2.0$.
This gives $\lambda=1.08$, the renormalized phonon frequency $\omega_{\rm r}\sim0.2$, and the inverse transition temperature $\beta_c=55$.
$\gamma_{\text {el}}$ is chosen to be much smaller than the SC gap in order not to break the SC order, while $\gamma_{\rm ph}$ is chosen so that the damping of phonons (the phonon peak width in the spectrum) mainly originates from the electron-phonon interaction.
We have also considered stronger $g$, different temperatures and both stronger and weaker $\gamma_{\rm ph}$ and $\gamma_{\rm el}$, and confirmed that the behavior discussed below does not change qualitatively.

 \begin{figure}[t]
  \centering
    \hspace{-0.cm}
    \vspace{0.0cm}
   \includegraphics{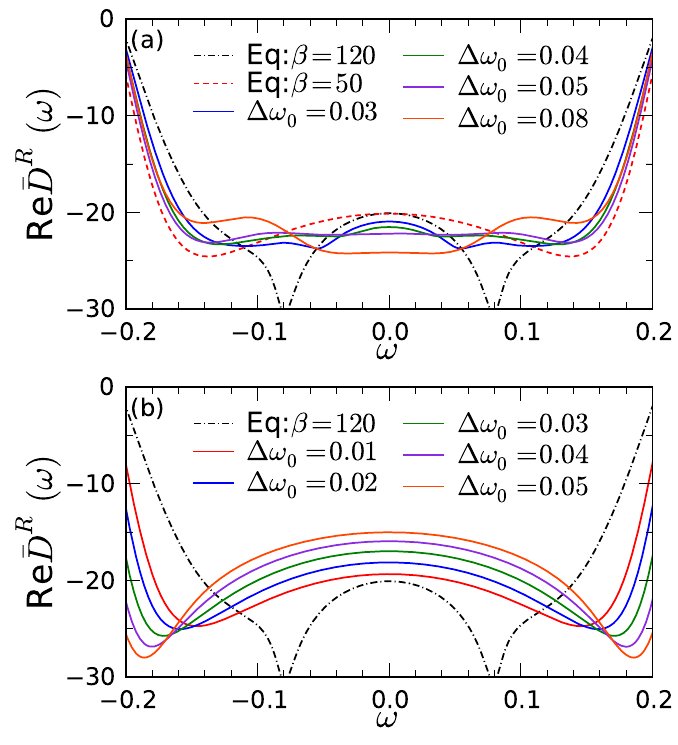}
  \caption{Real part of the retarded and time-averaged full phonon propagator, ${\rm Re}\bar{D}^R(\omega)$, for $g=0.41,\omega_0=0.4,\beta=120$ and the bath parameters $\gamma_{\rm ph}=0.02,\omega_D=0.6,\gamma_{\text {el}}=0.005,W_{\rm ebath}=2.0$. The equilibrium phonon propagator in the normal phase at $\beta=50$ is also shown in (a) for comparison. (a) is for $\Omega=0.125$ and (b) is for $\Omega=0.4$.}
  \label{fig:FDMFT_DR_Knap}
\end{figure}
\begin{figure}[t]
  \centering
    \hspace{-0.0cm}
    \vspace{0.0cm}
   \includegraphics{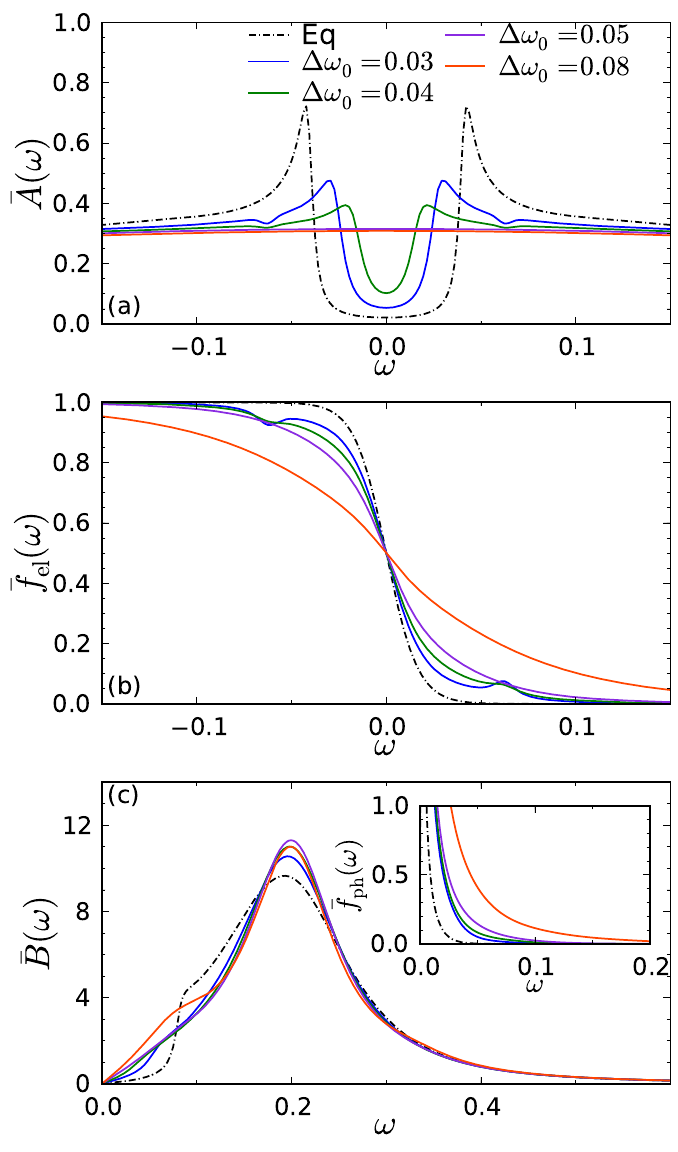}
  \caption{Time-averaged spectrum and distribution functions for $g=0.41,\omega_0=0.4,\beta=120$ and the bath parameters $\gamma_{\rm ph}=0.02,\omega_D=0.6,\gamma_{\text {el}}=0.005,W_{\rm ebath}=2.0$ and the excitation frequency $\Omega=0.125$. (a) Time-averaged electron spectrum. (b) Time-averaged electron distribution function. (c) Time-averaged phonon spectrum. The inset shows the time-averaged phonon distribution function.}
  \label{fig:FDMFT_spectrums_Knap}
\end{figure}

First, in Fig.~\ref{fig:FDMFT_DR_Knap}, we show $\text{Re}\bar{D}^R(\omega)$, which includes the effects of  $\Omega$ and $\Delta \omega_0$ as well as the renormalization through the electron-phonon coupling. 
In equilibrium, because of the strong el-ph coupling, the sign change in $\text{Re}\bar{D}^R(\omega)$ around the renormalized phonon frequency, $\omega_r$, is smeared out, see $\beta=50$ (above $T_c$) in Fig.~\ref{fig:FDMFT_DR_Knap}(a). Below $T_c$ [$\beta=120$ in Fig.~\ref{fig:FDMFT_DR_Knap}(a)], even though $|\bar{D}^R(0)|$ remains almost unchanged, a strong renormalization occurs for $\omega\neq0$.
This is caused by the opening of the SC gap, within which the scattering between electrons and phonons is suppressed. 
The peak in $|\text{Re}\bar{D}^R(\omega)|$ around $\omega=0.08$ corresponds to this energy scale.
This effect is known as the phonon anomaly in strongly coupled el-ph superconductors,\cite{Axe1973,Allen1997,Weber2008,Murakami2016} see also Fig.~\ref{fig:FDMFT_spectrums_Knap}(c). 
In the NESS, as in the case of the bare phonon propagator, the sign of the change in the effective attractive interaction $g^2\bar{D}^R(0)$ depends on the excitation frequency $\Omega$, despite the strong renormalization and the absence of sharp structures. 
Namely, for $\Omega\lesssim\omega_{\rm r}$, the strength $|g^2\bar{D}^R(0)|$ increases [see Fig.~\ref{fig:FDMFT_DR_Knap}(a)], while for $\Omega\gtrsim\omega_{\rm r}$, it decreases [see Fig.~\ref{fig:FDMFT_DR_Knap}(b)]. 
However, this enhancement of $D$ turns out not to lead to an enhancement of the superconductivity. 

Figure~\ref{fig:FDMFT_spectrums_Knap} illustrates how the (time-averaged) nonequilibrium spectrum and the distribution functions for the electrons and phonons change as we increase
the amplitude of the excitation. Here we choose $\Omega=0.125$ for which an enhancement of the effective attractive interaction has been observed [Fig.~\ref{fig:FDMFT_DR_Knap}(a)].
The nonequilibrium spectrum is defined as $\bar{A}(\omega)\equiv-\frac{1}{\pi}{\rm Im} \bar{G}^R(\omega)$ and the distribution function is $\bar{f}_{\text {el}}(\omega)\equiv \bar{N}(\omega)/\bar{A}(\omega)$ with $\bar{N}(\omega)=\frac{1}{2\pi} {\rm Im} \bar{G}^<(\omega)$.
 Here $\bar{G}^{R(<)}(\omega)=\frac{1}{\mathscr{T}}\int_0^\mathscr{T} dt_{\rm av} \int^\infty_{-\infty} dt_{\rm r} G^{R(<)}(t_{\rm r};t_{\rm av})e^{i\omega t_{\rm r}}$.  
As we increase the excitation amplitude $\Delta\omega_0$, the size of the SC gap decreases and eventually closes completely, see Fig.~\ref{fig:FDMFT_spectrums_Knap}(a).
At the same time, the slope of the nonthermal distribution function for the electrons decreases around $\omega=0$ and high energy states become more occupied, which roughly resembles the results for equilibrium states with higher temperatures, see Fig.~\ref{fig:FDMFT_spectrums_Knap}(b).

 \begin{figure}[t]
  \centering
        \hspace{-0.6cm}
    \vspace{0.0cm}
   \includegraphics{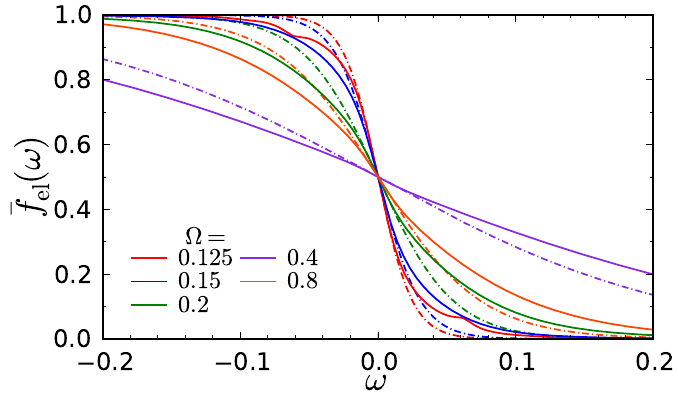}
  \caption{Time-averaged distribution functions (solid lines) for $g=0.41,\omega_0=0.4,\beta=120$ and the bath parameters $\gamma_{\rm ph}=0.02,\omega_D=0.6,\gamma_{\text {el}}=0.005,W_{\rm ebath}=2.0$ and $\Delta \omega_0=0.04$. 
  Dashed lines show the corresponding equilibrium distribution functions with the effective temperatures estimated from the slope $\partial_\omega \bar{f}_{\text {el}}(\omega)|_{\omega=0}$.}
  \label{fig:FDMFT_noneq_distribution_dum}
\end{figure}

One interesting feature in the nonequilibrium SC before melting is that there emerges a dip in $\bar{A}(\omega)$ and a hump in $\bar{f}_{\text {el}}(\omega)$ at $\omega=\Omega/2$. 
These features are only pronounced in the NESS with SC and disappear in the normal state. 
They can be explained as follows as a consequence of the two-band structure in the SC phase.  
The hump in $\bar{f}_{\text {el}}(\omega)$ results from excitations from the lower band of the Bogoliubov quasiparticles ($-E_{\rm k}$) to the upper band ($E_{\rm k}$), which  become efficient at $E_{\rm k}=\frac{\Omega}{2}$, where $E_{\rm k}$ is the energy of the quasiparticles. 
Although Bogoliubov excitations can recombine or relax within the bands due to the scattering with phonons and  the coupling to the environment, the continuous creation of particles at the particular energy $E_{\rm k}=\frac{\Omega}{2}$  can lead to an enhanced distribution in the steady state at the corresponding energy.
The dip in the spectral function can be explained as a hybridization effect between the first Floquet sideband of the upper (lower) Bogoliubov band with the lower (upper) Bogoliubov band, which occurs at $E_{\rm k}=\frac{\Omega}{2}$. The hybridization opens a gap at the crossing point, which is not fully developed in our case because of the finite correlations, resulting in a dip in the spectrum.

The phonon spectra, which are defined in the same manner as the electron spectra using $D$, show a hardening of the renormalized mode as we increase the field strength [Fig.~\ref{fig:FDMFT_spectrums_Knap}(c)], again consistent with the behavior seen in equilibrium with increased temperature.\cite{Murakami2015}
At the same time, the phonon distribution function defined as $\bar{f}_{\rm ph}(\omega)\equiv -\frac{1}{2\pi} {\rm Im} \bar{D}^<(\omega) /\bar{B}(\omega)$ shows an enhancement of the phonon occupation [see the inset in Fig.~\ref{fig:FDMFT_spectrums_Knap}(c)].
When we further increase the excitation strength, the Floquet sidebands of the phonons become more prominent [see $\Delta \omega=0.08$ data in Fig.~\ref{fig:FDMFT_spectrums_Knap}(c)],
in which case the spectrum does not resemble that at elevated temperatures.

Next, let us discuss how the nonthermal distribution is different from a thermal one. 
In Fig.~\ref{fig:FDMFT_noneq_distribution_dum},
we show the comparison between the nonthermal distribution and the thermal distribution 
with an effective temperature derived from $\beta_{\text {eff}}=-4\partial_\omega \bar{f}_{\text {el}}(\omega)|_{\omega=0}$.
We can see that this thermal distribution always underestimates the nonequilibrium distribution for $\omega>0$.
Therefore, it is expected that superconductivity will be weaker than what is expected from this effective temperature,
which is indeed the case, as shown below.

 \begin{figure}[t]
  \centering
      \hspace{-0.6cm}
    \vspace{0.0cm}
   \includegraphics[width=86mm]{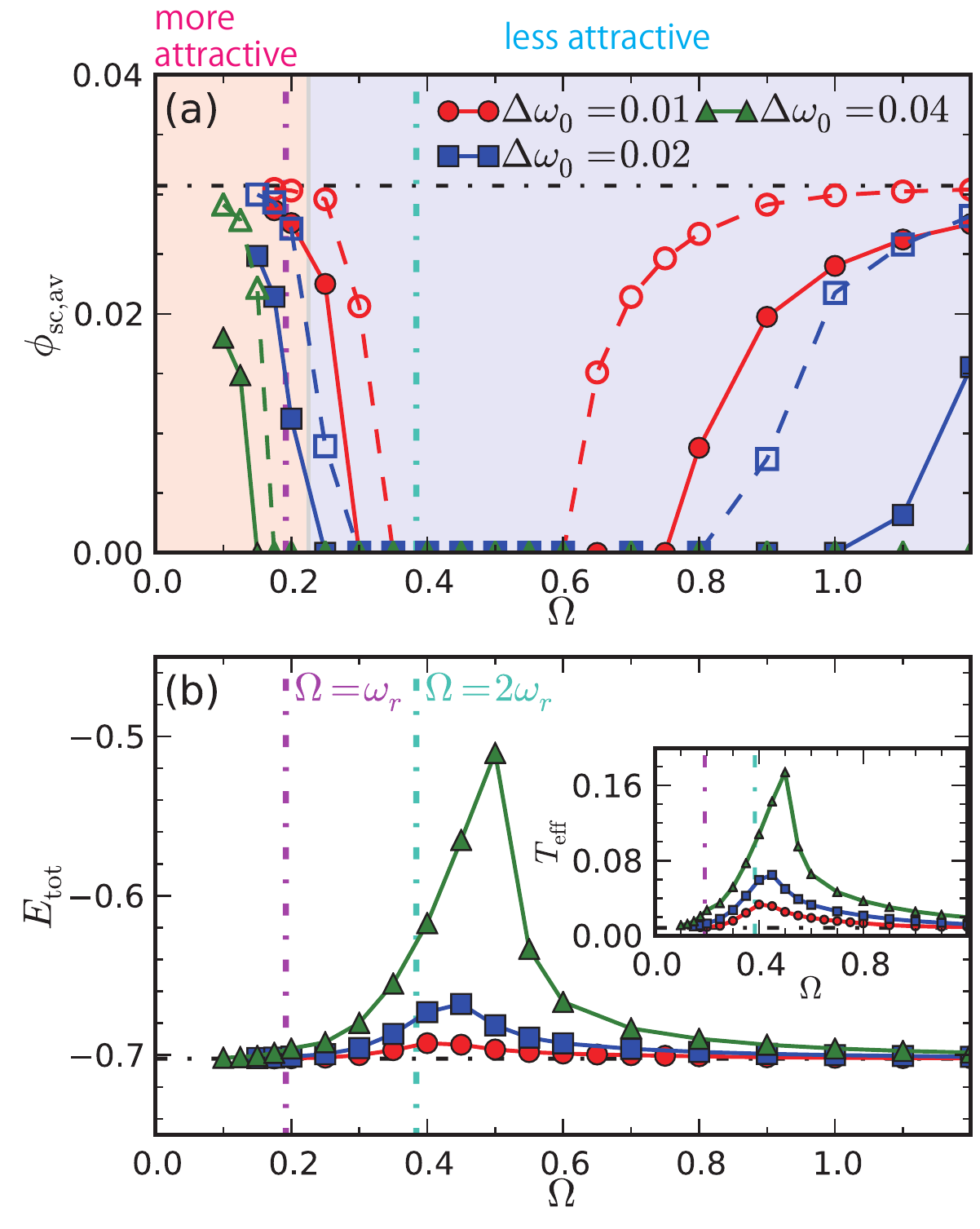}
  \caption{Time-averaged results for $g=0.41,\omega_0=0.4,\beta=120$ and the bath parameters $\gamma_{\rm ph}=0.02,\omega_D=0.6,\gamma_{\text {el}}=0.005,W_{\rm ebath}=2.0$. (a) Dependence of the SC order parameter $\phi_{\rm sc,av}$ on $\Omega$ for some $\Delta \omega_0$. The open symbols show the results estimated from the effective temperature. In the pink (blue) background region, $|\bar{D}^R(0)|$ increases (decreases) when $\Delta \omega_0$ changes from $0$ to $0.02$.
  (b) The dependence of $E_{\rm tot,av}$ on $\Omega$ for some $\Delta \omega_0$. The inset is the effective temperature estimated from the slope $\partial_\omega \bar{f}_{\text {el}}(\omega)|_{\omega=0}$. 
  The vertical lines indicate the positions of the parametric resonance in the weakly driven limit ($\Omega=\omega_{\rm r}, 2\omega_{\rm r}$).} 
  \label{fig:FDMFT_summary_dum}
\end{figure}

In Fig.~\ref{fig:FDMFT_summary_dum},
 we summarize our representative Floquet DMFT results in the strong el-ph coupling regime for quantities time-averaged over one period.
The SC order parameter is generally reduced from the equilibrium value (in fact, despite an extensive search, we could not find any parameter set where the SC order is enhanced in the strong-coupling regime, $\lambda\sim1$).
In particular, around $\Omega=0.4$, which is almost twice the renormalized phonon frequency, the superconducting order vanishes even for phonon modulations with small amplitude $\Delta \omega_0$.
This is consistent with the amount of energy injected by the excitation, and the effective temperature 
estimated from the nonthermal distribution function $\partial_\omega \bar{f}_{\text {el}}(\omega)|_{\omega=0}$, see Fig.~\ref{fig:FDMFT_summary_dum}(b).
We note that the peak position in Fig.~\ref{fig:FDMFT_summary_dum}(b) slightly shifts to higher energy as we increase the excitation strength.
This is because the injected energy changes the renormalization of the phonons, which affects the value of the parametric resonance ($\sim2\omega_{\rm r}$).
We also note that the SC order parameter is always lower than the equilibrium value at $T_{\text {eff}}$. 
This is consistent with the fact that the equilibrium distribution function at $T_{\text {eff}}$ underestimates the occupancy of the states at $\omega>0$.

Let us briefly comment on the effect of stronger couplings to heat baths, since one expects in this case more efficient energy absorption.
We have checked that stronger couplings to the phonon bath lead to a broader phonon peak in the spectrum, which results in a stronger suppression of SC and higher effective temperatures at $\Omega\lesssim \omega_r$.
Stronger couplings to the electron bath generally lead to a lower effective temperature and a less dramatic suppression of SC.
However, the equilibrium value of the SC order becomes smaller and in any case we only find a suppression of SC as far as we have investigated.

 \begin{figure}[t]
  \centering
    \hspace{-0.6cm}
    \vspace{0.0cm}
   \includegraphics[width=90mm]{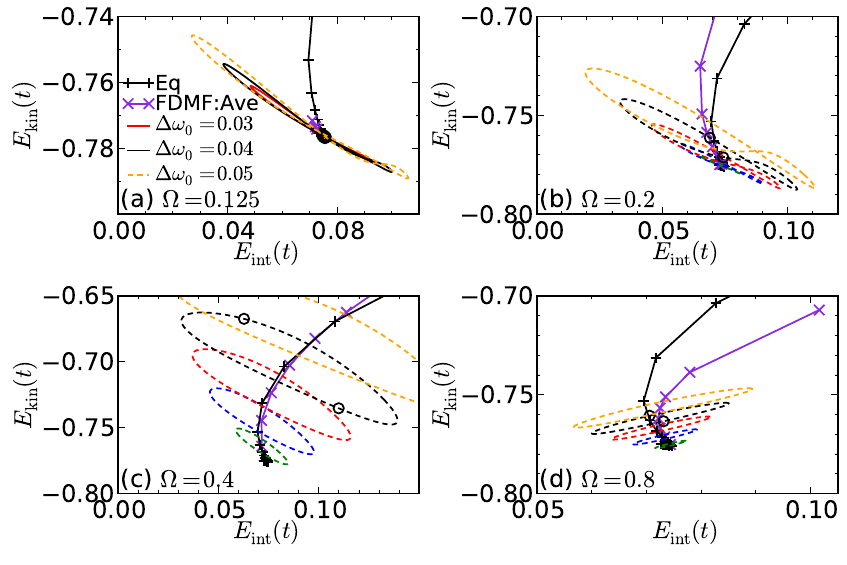}
  \caption{Trajectories of $E_{\rm kin}(t)$ and  $E_{\rm int}(t)=E_{\rm nX}(t)+E_{\rm ph}(t)$ in the NESS for $g=0.41,\omega_0=0.4,\beta=120$ and the bath parameters $\gamma_{\rm ph}=0.02,\omega_D=0.6,\gamma_{\text {el}}=0.005,W_{\rm ebath}=2.0$. Solid loops indicate that the NESS is in the SC state, while dashed loops are for the normal state. 
  In addition, we show the time-averaged values $\bar{E}_{\rm kin}$ and $\bar{E}_{\rm int}$ of the NESS for different excitation strength (violet curves), and the equilibrium $E_{\rm kin}$ and $E_{\rm int}$ for different temperatures (black lines).
  Black open circles indicate the values at $t=\frac{\pi}{2\Omega},\frac{3\pi}{2\Omega}$, where the Hamiltonian temporarily becomes the same as the non-perturbed one for $\Delta \omega=0.04$.}
  \label{fig:Ekin_vs_Eint_Knap}
\end{figure}
So far we have focused on the time-averaged values, whereas in the NESS all the quantities are oscillating around the averaged value during one period. 
In Fig.~\ref{fig:Ekin_vs_Eint_Knap}, we plot the trajectories of $E_{\rm kin}(t)$ and  $E_{\rm int}(t)=E_{\rm nX}(t)+E_{\rm ph}(t)$ for the NESS over one period, and compare them to the averaged $\bar{E}_{\rm kin}$ and $\bar{E}_{\rm int}$ of the NESS (violet lines) and the equilibrium $E_{\rm kin}$ and $E_{\rm int}$ for different temperatures (black lines).
First, we can see that the results for the time averaged values of $E_{\rm kin}$ and $E_{\rm int}$ are in general not on top of the equilibrium curve, 
and that the deviation strongly depends on the excitation frequency. This indicates that it is difficult to fully describe the properties of NESSs through an analogy to  equilibrium states with elevated temperatures. 
We also notice that the oscillations around the averaged values are large, and that they become largest at the parametric resonance, $\Omega=0.4\simeq 2\omega_r$. Even at the times where the Hamiltonian becomes temporarily identical to the unperturbed one ($t=\frac{\pi}{2\Omega},\frac{3\pi}{2\Omega}$) the energies can strongly deviate from the thermal line. This indicates that the dynamics which occurs during one period cannot be captured using an adiabatic picture either in these cases.

 \begin{figure}[t]
  \centering
    \hspace{-0.6cm}
    \vspace{0.0cm}
   \includegraphics[width=90mm]{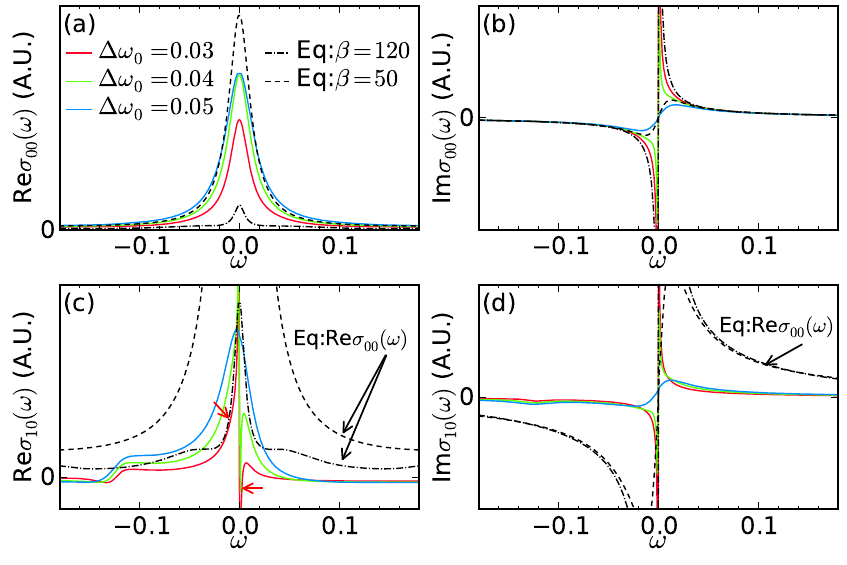} 
  \caption{Optical conductivity ($\sigma_{m0}(\omega)$) for $g=0.41,\omega_0=0.4,\beta=120$ and the bath parameters $\gamma_{\rm ph}=0.02,\omega_D=0.6,\gamma_{\text {el}}=0.005,W_{\rm ebath}=2.0$ and the excitation $\Omega=0.125$. $\Delta\omega_0=0.03,0.04$ and the equilibrium state at $\beta=120$ are in SC, and the rest are in the normal phase. (a)(b) The diagonal component of the optical conductivity, $\sigma_{00}(\omega)$. (c)(d) The off-diagonal component of the optical conductivity, $\sigma_{10}(\omega)$. The dotted line and the dashed and dot-dashed lines show the equilibrium $\sigma_{00}(\omega)$ with indicated temperatures for comparison. The red allows in (c) point to the $1/\omega$ divergence for $\Delta\omega_0=0.03$. At $\omega=0$ in the SC states, the delta function is located at each $m$, which is not shown here.}
  \label{fig:ness_conductivity}
\end{figure}

In Fig.~\ref{fig:ness_conductivity}, we show the optical conductivity in the NESS.
In equilibrium, for $T<T_c$, the real part shows a reduction of the spectral weight below the SC gap 
 and the imaginary part exhibits a $1/\omega$ dependence, whose coefficient corresponds to the superfluid density. 
In the NESS, the optical conductivity $\sigma(t,t')$, like any correlation function, can be written in the Floquet matrix form Eq.\eqref{Fmatrix111}, see the discussion around Eq.~(\ref{eq:cond}).
In the diagonal component in the NESSs, 
the weight of the real part inside the SC gap 
 increases with increasing field strength,  
 and becomes close to the result at elevated temperatures [see Fig.~\ref{fig:ness_conductivity}(a)].
 Meanwhile, the $1/\omega$ component in the imaginary part is suppressed, which indicates the suppression of the superfluid density and is consistent with the 
closure of the gap and the decrease of the SC order parameter [see Fig.~\ref{fig:ness_conductivity}(b)]. 
In the off-diagonal component, one can also observe a $1/\omega$ around $\omega=0$ behavior both in the real and imaginary parts in the SC phase [see Fig.~\ref{fig:ness_conductivity}(c)(d)].
(We note that $\sigma_{m,n}(\omega)^*=\sigma_{-m,-n}(-\omega)$.)
This is caused from the imperfect cancellation between $\chi^{\rm para}$ and $\chi^{\rm dia}$, and corresponds to the following asymptotic form for $|t-t'| \rightarrow \infty$ of the conductivity,
\begin{align}
\sigma_S (t,t')\equiv\theta(t-t') \sum_m n_{S,m} e^{-im \Omega t}.
\end{align}
We note that $n_{S,m}$ can be a complex number and $n_{S,m}^*=n_{S,-m}$. 
This indicates that if a delta-function electric-field pulse is applied to the system, the induced current continues to flow forever  (supercurrent) oscillating with a period of $2\pi/\Omega$,
and the phase of $n_{S,m}$ determines the phase of the oscillations in the suppercurrent.
 The $1/\omega$ component in $\sigma_{m0}(\omega)$ corresponds to $n_{S,m}$.
For example, without the phonon driving $n_{S,0}\approx 1.18$, while for $\Delta\omega_0=0.03$ $n_{S,0}\approx 0.62$ and $n_{S,1}\approx 0.05-i0.01$.


\subsection{Transient dynamics towards the NESS}\label{sec:Trans1}
In order to study the transient dynamics towards the NESS, 
we consider a situation where the system at time $t=0$ is in equilibrium (attached to the bath) and at $t>0$
is exposed to a phonon driving of type 1 with 
\begin{align}
\omega_{0\rm X}(t)=\omega_0+\Delta\omega_0 \sin(\Omega t). \label{eq:excitation_trans}
\end{align}
In order to follow the transient dynamics, we use the nonequilibrium DMFT formulated on the Kadanoff-Baym contour.\cite{Aoki2013,Murakami2015,Murakami2016}
For this, we first need to evaluate the bare phonon propagator subject to the periodic driving by
solving Eq.~(\ref{eq:D0_eq}). Apart from that, the calculation is identical to the scheme explained in Refs.~\onlinecite{Murakami2015,Murakami2016}.

 \begin{figure}[t]
  \centering
  \hspace{-0.6cm}
    \vspace{0.0cm}
   \includegraphics[width=90mm]{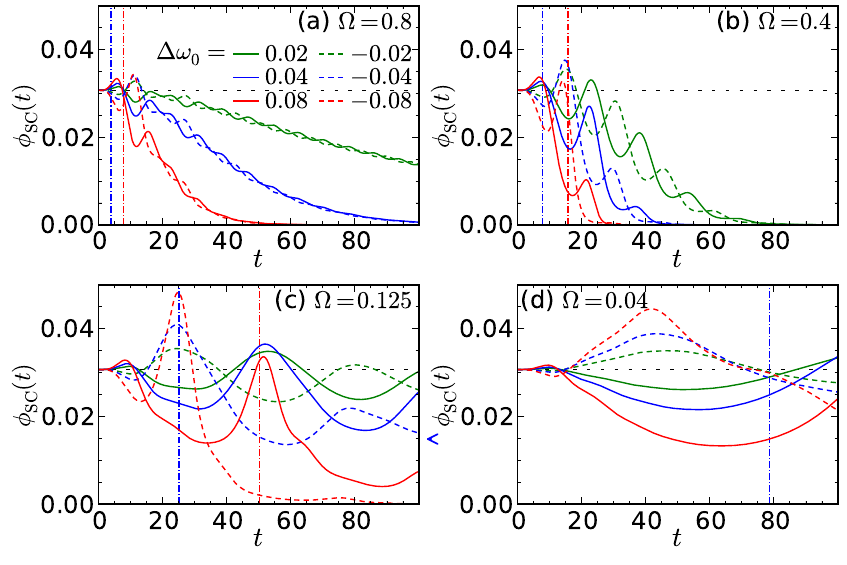} 
  \caption{Time evolution of the SC order parameter under the periodic driving of type 1 for various frequencies and amplitudes for $g=0.41,\omega_0=0.4,\beta=120$ and the bath parameters $\gamma_{\rm ph}=0.02,\omega_D=0.6,\gamma_{\text {el}}=0.005,W_{\rm ebath}=2.0$. 
  Horizontal black dotted lines indicate the equilibrium value, while the blue and red vertical lines show $t=\pi/\Omega$ and $t=2\pi/\Omega$, respectively. The arrow in panel (c) indicates the value at the NESS.}
  \label{fig:phi_knap1}
\end{figure}
In Fig.~\ref{fig:phi_knap1}, we show the transient dynamics of the SC order parameter, $\phi_{\rm SC}(t)$, for various excitation conditions.
First we note that when $\omega_{0\rm X}$ becomes small with $g$ and $T(<T_c)$ fixed, $\phi_{\rm SC}$ becomes larger because the system moves to the stronger coupling regime.
This leads to the expectation that in the adiabatic regime, $\phi_{\rm SC}$ becomes larger with decreasing $\omega_{0\rm X}$. 
In other words, for $\Delta \omega_0>0$ (solid lines in Fig.~\ref{fig:phi_knap1}) $\phi_{\rm SC}$ should initially decrease, while for $\Delta\omega_0<0$ (dashed lines) it should increase. 
However, in the initial dynamics for non-adiabatic excitations, the solid (dashed) lines show a positive (negative) hump, which is opposite to the above expectation.
The hump becomes smaller with decreasing $\Omega$, and the dynamics becomes more consistent with the expectations from the adiabatic picture.
We also note that, during the first period, one can see a significant transient enhancement of the order parameter in some cases, see Fig.~\ref{fig:phi_knap1}. 
There the time-averaged $\phi_{\rm SC}(t)$ for one period can become larger than the equilibrium value if we carefully choose the phase of the excitation, i.e. the sign of $\Delta \omega_0$.

As time evolves further, $\phi_{\rm SC}(t)$ is however strongly suppressed from the initial value.
One can see that the fastest suppression occurs around $\Omega=0.4\sim2\omega_{\rm r}$, which is at the 
parametric resonance.
This is consistent with the analysis of the NESS with the Floquet DMFT, where we have observed the 
strongest energy injection and the strongest increase of the effective temperature in this driving regime.
The results in Fig.~\ref{fig:phi_knap1} show that even in the transient dynamics, the parametric resonance is associated with a rapid destruction of the SC order.

 \begin{figure}[t]
  \centering
  \hspace{-0.6cm}
    \vspace{0.0cm}
   \includegraphics[width=90mm]{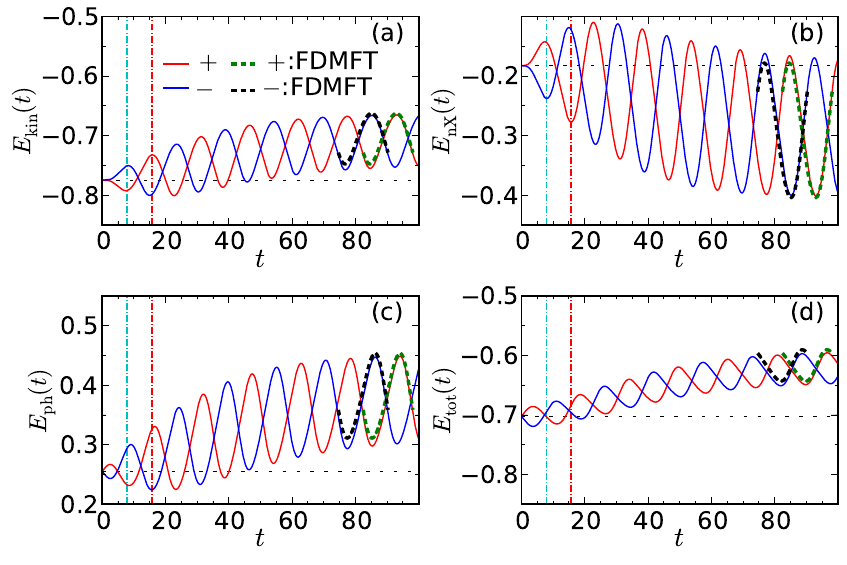} 
  \caption{Time evolution of the energies for the type 1 excitation with $\Omega=0.4$ and $|\Delta \omega_0|=0.04$ for $g=0.41,\omega_0=0.4,\beta=120$ and the bath parameters $\gamma_{\rm ph}=0.02,\omega_D=0.6,\gamma_{\text {el}}=0.005,W_{\rm ebath}=2.0$. The sign of $\Delta \omega_0$ is indicated in the figures.  Horizontal black dotted lines indicate the equilibrium value, while the blue and red vertical lines show $t=\pi/\Omega$ and $t=2\pi/\Omega$, respectively.}
  \label{fig:en_knap1}
\end{figure}
In Fig.~\ref{fig:en_knap1}, we show how the energies approach the NESS described by the Floquet DMFT.
These results demonstrate that it takes several cycles for the system to reach the NESS.

 \begin{figure}[t]
  \centering
  \hspace{-0.6cm}
    \vspace{0.0cm}
   \includegraphics{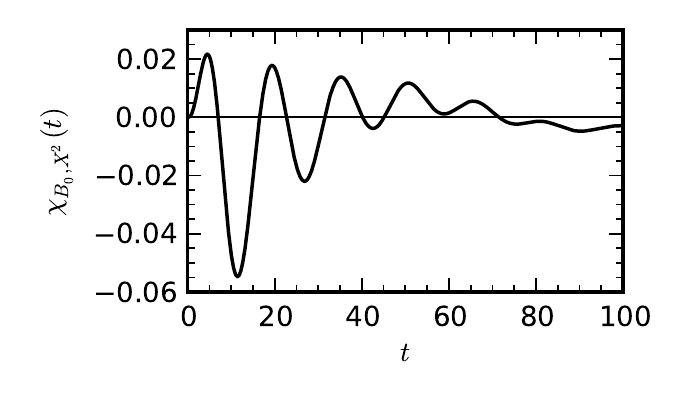} 
  \caption{$\chi_{B_{\bf 0},X^2} (t)$ for $g=0.41,\omega_0=0.4,\beta=120$ and the bath parameters $\gamma_{\rm ph}=0.02,\omega_D=0.6,\gamma_{\text {el}}=0.005,W_{\rm ebath}=2.0$.}
  \label{fig:chi_BX}
\end{figure}

 \begin{figure}[t]
  \centering
  \hspace{-0.6cm}
    \vspace{0.0cm}
   \includegraphics[width=90mm]{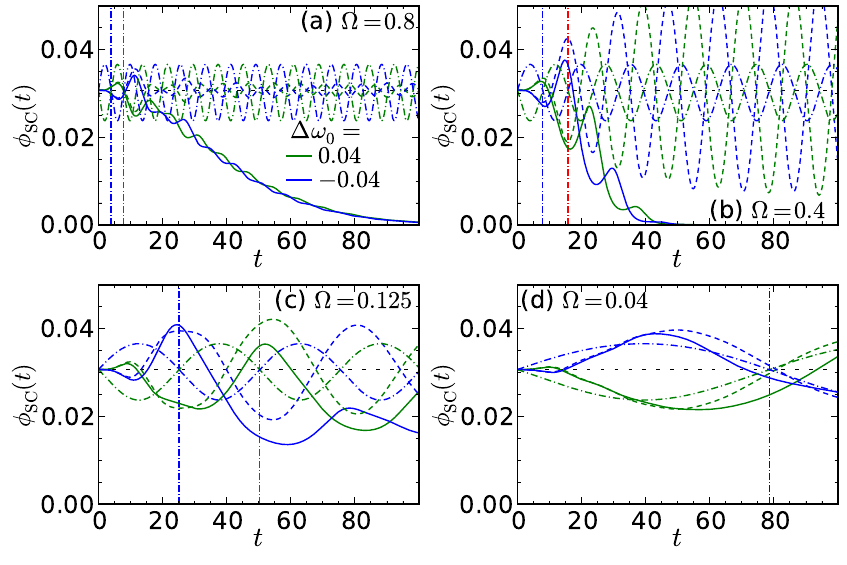} 
  \caption{Comparison between the full dynamics of the SC order parameter (solid lines) and the prediction from the linear-response theory (dashed lines) for $g=0.41,\omega_0=0.4,\beta=120$ with the bath $\gamma_{\rm ph}=0.02,\omega_D=0.6,\gamma_{\text {el}}=0.005,W_{\rm ebath}=2.0$. The sign  of $\Delta \omega_0$ is indicated in the figures. Horizontal black dotted lines indicate the equilibrium values, while the blue and red vertical lines show $t=\pi/\Omega$ and $t=2\pi/\Omega$, respectively. Dash-dotted curves show the results of the adiabatic approximation for the time-dependent Hamiltonian at $\beta=120$.}
  \label{fig:time_evolution_linear_response}
\end{figure}
To gain more insights into the dynamics of the order parameter, we compare it with the linear-response result. In the linear-response regime, the variation of the order parameter can be expressed as 
\begin{align}\label{eq:chi_BXX}
\Delta B_{\bf 0}(t)&=\int d\bar{t} \chi_{B_{\bf 0},X^2} (t-\bar{t}) F(\bar{t}), 
\end{align}
with $\chi_{B_{\bf 0},X^2} (t)\equiv -i\theta(t) \langle [B_{\bf 0}(t),X^2(0)]\rangle$, $B_{\bf 0}=\sum_i c^\dagger_{i\uparrow}c^\dagger_{i\downarrow}+H.c.$, and $F(t)=\Delta\omega_0\sin(\Omega t)$. 
The linear-response function can be numerically evaluated by considering, instead of the periodic driving, $H_{\rm ex}(t)=d_f\delta(t) \sum_i X_i^2$ and taking $d_f$ small enough.
The resulting response function is shown in Fig.~\ref{fig:chi_BX}.
The oscillation frequency in $\chi_{B_{\bf 0},X^2} (t)$ is $\omega\sim 0.4\sim2\omega_{\rm r}$.
\footnote{With the phonon bath it turns out that the phonon renormalization and the effective phonon-phonon interaction discussed in Ref.~\onlinecite{Murakami2016} become weaker than without. Hence the oscillation frequency in $\chi_{B_{\bf 0},X^2}$ remains close to $2\omega_{\rm r}$ in the present case.}.

In Fig.~\ref{fig:time_evolution_linear_response}, we compare the full dynamics (solid lines) and the linear-response component (dashed lines), which is obtained by taking the convolution, Eq.~(\ref{eq:chi_BXX}).
The difference corresponds to the contributions from the non-linear components.
The linear-response theory captures the initial behavior of the order parameter including the temporal enhancement or suppression.
The initial hump whose direction is opposite to the adiabatic expectation is also captured.
This hump originates from the positive value of $\chi_{B_{\bf 0},X^2} (t)$ just after $t=0$.
Since the first derivative of $\chi_{B_{\bf 0},X^2} (t)$ at $t=0$ is zero, as can be analytically shown, 
this corresponds to the positive value of $\partial^2_t\chi_{B_{\bf 0},X^2} (t)|_{t=0}=8ig\omega_0\sum_i\langle(-c^\dagger_{i\uparrow}c^\dagger_{i\downarrow}+c_{i\downarrow}c_{i\uparrow})X_i\rangle$. 
If we decrease $\Omega$ with $\Delta \omega_0$ fixed (as we approach the adiabatic regime), the contribution from the initial positive hump in $\chi_{B_{\bf 0},X^2} (t)$ decreases, and 
the hump in $\phi_{\rm SC}$ becomes smaller.
At longer times, $\phi_{\rm SC}(t)$ starts to deviate from the linear-response result, which generally overestimates $\phi_{\rm SC}(t)$.

This indicates that the general effect of the non-linear component is to reduce the superconducting order.
It is also illustrative to compare the results to the expectation in the adiabatic limit, in which the temperature is fixed to the bath temperature and the phonon potential is changed (see the dash-dotted curves in Fig.~\ref{fig:time_evolution_linear_response}). (Note that in an isolated system, the adiabatic limit would be defined by constant entropy instead of constant temperature.) The comparison confirms that the dynamics approaches the adiabatic expectation as we decrease the excitation frequency.


\subsection{Superconducting fluctuations above $T_c$ under phonon driving}\label{sec:Trans2}
So far we have focused on the superconducting phase in the strong-coupling regime and showed that the 
phonon driving generally weakens the superconductivity and that this tendency is particularly pronounced in the parametric resonance regime.
Now we move on to the weak-coupling regime ($\lambda \ll 1$), where the BCS theory has usually been applied.
In this regime the transition temperature becomes very small and a direct investigation of the superconducting state is numerically difficult.
Hence we study the time evolution of the superconducting fluctuations in the normal state, in systems with and without 
phonon driving. 
Specifically, we evaluate the dynamical pair susceptibility under the excitation described by Eq.~(\ref{eq:excitation_trans}).
For this, as in the equilibrium case,\cite{Murakami2016} we add $H'_{\rm ex}(t)=F_{\rm ex}(t)B_{\bf 0}$ on top of the phonon modulation, where $B_{\bf 0}=\sum_i(c_{i\uparrow}^\dagger c_{i\downarrow}^\dagger+c_{i\downarrow}c_{i\uparrow})$ and $F_{\rm ex}(t)=d_{\rm f}\delta(t)$.  We take $d_{\rm f}$ small enough and compute the time evolution of  $B_{\bf 0}$ to obtain the dynamical pair susceptibility,
\begin{align}
 \chi^R_{\rm pair}(t,t')=-i\theta(t-t') \frac{1}{N} \langle [B_{\bf 0}(t),B_{\bf 0}(t')] \rangle.
 \end{align} 
 In the normal state in equilibrium, this susceptibility decays as $t-t'$ increases. The decay time increases as we approach the boundary to the superconducting state and diverges at the boundary.
 The interesting question thus is how this decay depends on the phonon driving in the weak-coupling regime.
 
 \begin{figure}[t]
  \centering
        \hspace{-0.6cm}
      \vspace{0.cm}
   \includegraphics[width=90mm]{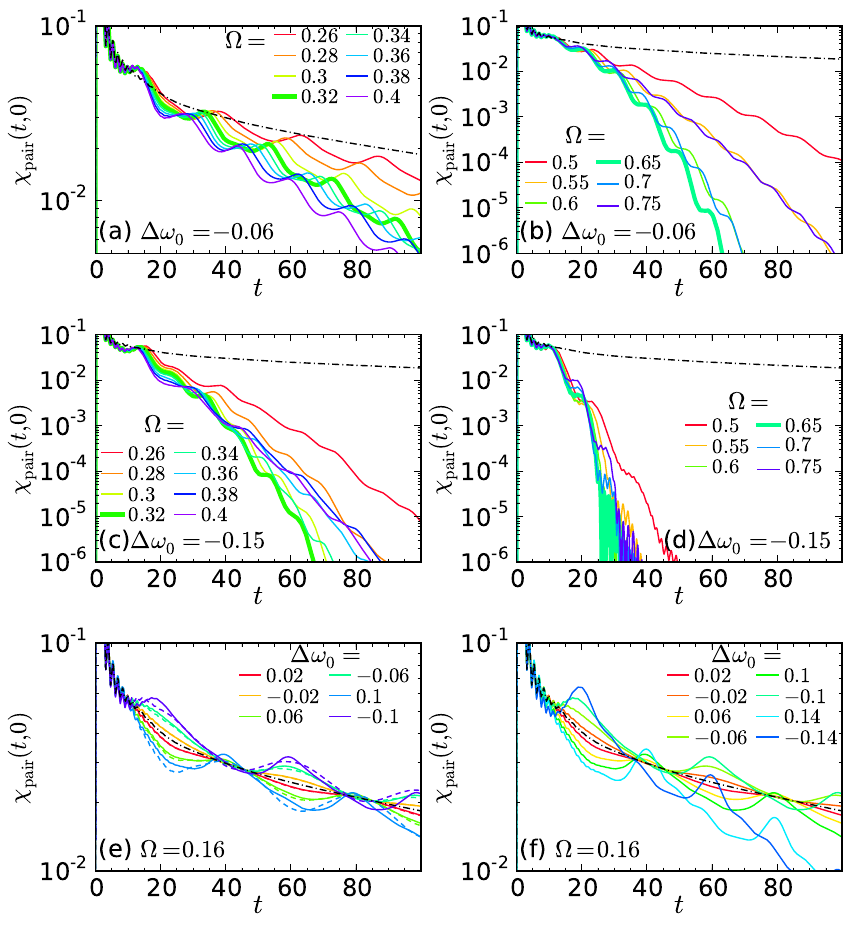}
  \caption{Transient dynamics of the SC fluctuations above $T_c$ under the periodic driving of type 1, $\chi_{\rm pair}(t,0)$. The parameters are $g=0.3,\omega_0=0.4,\beta=320$ and no heat bath is attached.  
 Panels (a)(c) are for driving frequencies around $\Omega=\omega_{\rm r}$, while (b)(d) are for frequencies around $\Omega=2\omega_{\rm r}$.
  The data at the resonance are shown by bold lines.
(e)(f) show the results for $\Omega=0.16$ and various $\Delta \omega_0$. 
  Dotted lines are the linear component (in $\Delta \omega_0$) estimated from the $\Delta \omega_0=\pm 0.02$ data.
  The dash-dotted black line in each panel is the equilibrium result.}
  \label{fig:sc_fluc_summary}
\end{figure}

\begin{figure}[t]
  \centering
        \hspace{-0.6cm}
      \vspace{0.cm}
   \includegraphics[width=90mm]{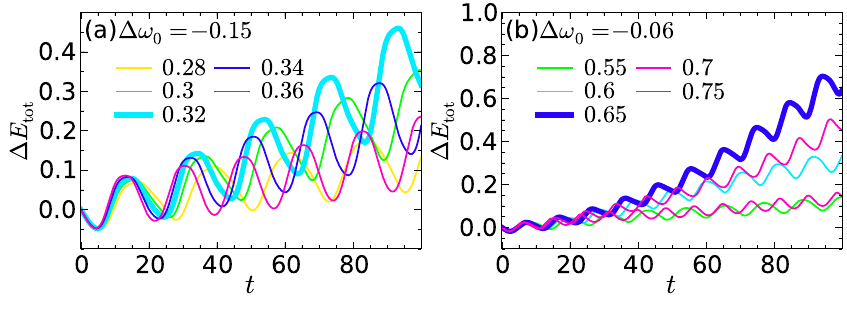}
  \caption{Time evolution of the total energy of the system under the periodic driving of type 1. The parameters are $g=0.3,\omega_0=0.4,\beta=320$ and no heat bath is attached.  
   Different colors indicate different $\Omega$. Panel (a) shows the results around $\Omega=\omega_{\rm r}$, while panel (b) is for the driving frequencies around $\Omega=2\omega_{\rm r}$. The data at the resonance are shown by bold lines.}
  \label{fig:energy_prof_g0.3gamel0.0b320}
\end{figure}

 In Fig.~\ref{fig:sc_fluc_summary}, we show the results for $g=0.3,\omega_0=0.4,\beta=320$ without heat baths.
 This condition corresponds to $\lambda\sim0.23$ and $\omega_{\rm r}=0.32$, and the inverse transition temperature is $\beta_c\sim1000$.
 By changing the temperature (down to $\beta=640$), $g$ and the couplings to the heat baths, we have confirmed that the behavior described below is the generic one in the weak-coupling regime. 
 In Fig.~\ref{fig:sc_fluc_summary}(a)(c), we show the results around $\Omega=\omega_{\rm r}$. 
 We remind the reader that from the analysis of $D_0$ in Sec.~\ref{sec:D0}, an strong enhancement of $D^R(0)$ is expected for $\Omega\lesssim\omega_{\rm r}$.
 However, what we observe is that generally in the presence of periodic phonon modulations the lifetime of the fluctuations becomes shorter than without phonon modulations
  and that at $\Omega=\omega_{\rm r}$ the lifetime is particularly strongly reduced. 
This tendency becomes clearer with increasing $\Delta\omega_0$.
 In Fig.~\ref{fig:energy_prof_g0.3gamel0.0b320}(a), we show the evolution of the energy absorption around the parametric resonance, $\Omega=\omega_{\rm r}$.
It turns out that at the parametric resonance the energy absorption is also very efficient.
These results suggest that even though there might be some enhancement of the attractive interaction, the energy absorption and resulting disturbance of long-range correlations is the dominant effect in a wide parameter range. In particular, at the parametric resonance, the 
maximized energy absorption rate 
leads to a particularly rapid heating of the system, and efficient destruction of the SC fluctuations. 
We also note that the decay of the fluctuations accelerates as $t$ increases, as is evidenced by the 
concave form of $ \chi^R_{\rm pair}(t,0)$ in the log scale. 
This can also be interpreted as a heating effect. The more energy the system gains as $t$ increases, the faster the decay of the fluctuation becomes.

This dominant heating effect and quick destruction of the SC fluctuation can also be observed in another resonant regime $\Omega\sim2\omega_{\rm r}$, where in the polaron picture an enhancement of the superconductivity is expected,\cite{Knap2016} see Fig.~\ref{fig:sc_fluc_summary}(b)(d).
Here, $\Omega=0.65\sim 2\omega_{\rm r}$ shows the faster decay than the neighboring driving frequencies. The efficient absorption of energy resulting from this parametric phonon driving is shown in Fig.~\ref{fig:energy_prof_g0.3gamel0.0b320}(b). 

Now, we point out that there is a regime where the phonon driving slightly enhances the SC fluctuations.
In Fig.~\ref{fig:sc_fluc_summary}(e)(f), we show the SC fluctuations for a smaller excitation frequency than those above, where less heating is expected, and for various excitation strengths.
In all cases, $ \chi^R_{\rm pair}(t,0)$ stays around the equilibrium curve, but exhibits a shift above or below the equilibrium value depending on the sign of $\Delta \omega_0$. 
For small amplitudes, the change of $ \chi^R_{\rm pair}(t,0)$ with respect to the equilibrium value $ \chi^R_{\rm pair,eq}(t)$ is linear in $\Delta \omega_0$. 
In order to identify the component linear in $\Delta \omega_0$, we rescale the difference of $ \chi^R_{\rm pair}(t,0)-\chi^R_{\rm pair,eq}(t)$ at the smallest amplitude $\Delta \omega_0=0.02$ to larger amplitudes, and depict the results with dashed lines in Fig.~\ref{fig:sc_fluc_summary}(e). 
The linear components consistently explain the general behavior. 
The enhancement of the SC fluctuations for $\Delta\omega_0<0$ can be explained as follows.
At $t\in[0,\pi/\Omega]\sim[0,20]$, the phonon frequency is temporarily reduced and the system is temporarily in a stronger coupling regime compared to equilibrium
(this is not the effect of the Floquet sidebands).
Hence the decay of the SC fluctuations is slower in this time interval. 
During $t\in[\pi/\Omega,2\pi/\Omega]$ the situation is opposite
and we observe a faster decay. 
However, because of the enhancement during $t\in[0,\pi/\Omega]$, $ \chi^R_{\rm pair}(t,0)$ stays above the equilibrium curve.
The opposite is the case when $\Delta\omega_0$ is positive.

The non-linear component can result in a slight but systematic increase of the SC fluctuations, in particular at $\Delta \omega_0=\pm0.06$, see the difference between the solid and dotted lines in Fig.~\ref{fig:sc_fluc_summary}(e). 
We note that both the heating and the creation of the Floquet sidebands of the phonons, which leads to the enhancement of the attractive interaction, start from the second order in the field strength. 
Therefore, this observation may indicate that the enhancement of the attractive interaction slightly dominates the heating effect.
However this enhancement is very subtle and as we further increase the field strength this enhancement becomes weak and eventually the SC fluctuations decay faster than in the equilibrium case, see Fig.~\ref{fig:sc_fluc_summary}(f).

Finally we note that even though we have also studied a parameter set  for $\omega_0>W$ in the weak coupling regime with various excitation conditions,
we could not observe any enhancement of the SC fluctuation.
\footnote{As far as the system is in the weak-coupling regime the Migdal approximation may still work.}

\section{Conclusions}\label{sec:conclusion}
We have systematically investigated the effects of the parametric phonon driving on the superconductivity (SC) using the simple Holstein model and the nonequilibrium dynamical mean-field theory (DMFT). In order to study the time-periodic nonequilibrium steady state, we extended the Floquet DMFT formalism, which has previously been applied to purely electronic systems, to this electron-phonon model. 
We have also studied the transient dynamics under periodic phonon driving with the nonequilibrium DMFT formulated on the Kadanoff-Baym contour. 
This simulation method, which does not involve a gradient approximation, can describe dynamics on arbitrarily fast timescales.

In the strong electron-phonon coupling regime, we have studied both the nonequilibrium steady states and the transient dynamics towards these states.
Even in the presence of the renormalization from the electron-phonon coupling, the effective attractive interaction characterized by $\bar D^R(\omega=0)$ shows an enhancement in some parameter regime.
However, taking into account heating and the nonthermal nature of the driven state, the ultimate effect of the driving turns out to be a suppression of the SC gap, the SC order parameter or the superfluid density both in the time-periodic steady states and the transient dynamics.
The strongest suppression of SC is observed in the parametric resonant regime, $\Omega\sim2\omega_r$. 

In the weak-coupling regime, we have studied the SC fluctuations above $T_c$.
We have found that, generally, they are suppressed by the phonon driving, in the sense that the pairing susceptibility decays more rapidly in the driven state. 
The decay becomes particularly fast in the parametric resonant regimes, which can be attributed to a particularly efficient energy absorption. 

From these analyses, we conclude that the 
previously predicted 
enhancement of the superconductivity in the parametric resonant regime is not a general result. In the simple Holstein model with a phonon energy scale smaller than that of electrons, such effects, if any, are overwhelmed by the absorption of energy
in a wide parameter range, and particularly strongly so near the parametric resonances. 
Even though our analysis focuses on a specific model, it demonstrates the importance of a careful treatment of heating effects and/or nonthermal distributions when discussing the potential effects of phonon driving.
We note that in addition to the coupling of light to phonons, one could also take into account the direct coupling of light to electrons, by means of a Peierls substitution. This can be expected to lead to further heating effects. 

As for the relation to the K$_3$C$_{60}$ experiment mentioned in the introduction,\cite{Mitrano2016} from our analysis it seems likely that the 
apparent enhancement of the SC originates from the static change of the Hamiltonian\cite{Kim2016,Mazza2017} or some other dynamical effects\cite{Kennes2017,Sentef2017,Nava2017}, which are not captured in the present simple model. Since alkali doped fullerides are multi-orbital systems and an effective negative Hund's coupling from the Jahn-Teller screening plays an important role,\cite{Capone2009,Nomura2015,Karim2016,Hoshino2016} we may need to go beyond the Holstein model to fully understand the mechanism.

Finally, we remark that the diagrammatic method for steady states and the transient dynamics used here can be easily extended to other scenarios for the enhancement of SC, such as non-linear couplings to electrons,\cite{Kennes2017,Sentef2017} 
 which is an interesting direction for future studies.
The investigation of nonequilibrium states with an unconventional pairing such as the d-wave superconductivity in cuprates is also an interesting future work. 
 In particular it is important to understand whether or not a particular mechanism for the enhancement of SC survives even if heating effects are taken into account.

\acknowledgments
 The authors wish to thank D. Gole$\check{\rm z}$, H. Strand and J. Okamoto for fruitful discussions.
YM is supported by the Swiss National Science Foundation through NCCR Marvel. 
PW acknowledges support from FP7 ERC Starting Grant No.~278023. NT is supported by JSPS KAKENHI (Grant No.~16K17729).
ME acknowledges support by the Deutsche Forschungsgemeinschaft within the Sonderforschungsbereich 925 (projects B4).

\appendix

\section{The expression for the energy dissipation in the time-periodic NESS}\label{sec:Harada}
The explicit expression for Eq.~\eqref{eq:Hrada_sasa} is
\begin{align}
\overline{I}&=\bar{I}_{\text {el}}+\bar{I}_{\rm ph}\nonumber\\
&=i\overline{\langle [H_{\rm ph\_mix}(t),H_{\rm ph\_bath}(t)]\rangle}+i\overline{\langle [H_{\rm el\_mix}(t),H_{\rm el\_bath}(t)]\rangle}.
\end{align}
Using the bath's fluctuation-dissipation relation,
we obtain
\begin{align}
\bar{I}_{\text {el}}&=-i\sum_{i,\sigma}\int^{\Omega/2}_{-\Omega/2}\frac{d\omega}{2\pi}\sum_n(\omega+n\Omega)\Gamma(\omega+n\Omega)\nonumber\\
&\left\{{\bf G}^K_{i,\sigma,nn}(\omega)-F_f(\omega+n\Omega)({\bf G}^R_{i,\sigma,nn}(\omega)-{\bf G}^A_{i,\sigma,nn}(\omega))\right\}, 
\end{align}
with $F_f(\omega)=\tanh(\frac{\beta\omega}{2})$ and
\begin{align}
\bar{I}_{\rm ph}&=\frac{i}{2} \sum_i\int^{\Omega/2}_{-\Omega/2} \frac{d\omega}{2\pi} \sum_n (\omega+n\Omega) B_{\rm bath}(\omega+n\Omega)\nonumber\\
&\left\{ {\bf D}_{i,nn}^K(\omega)-F_b(\omega+n\Omega)({\bf D}_{i,nn}^R(\omega)-{\bf D}_{i,nn}^A(\omega))\right\},
\end{align}
with $F_b(\omega)=\coth(\frac{\beta\omega}{2})$.

These expressions imply that the energy dissipation $\bar{I}$ is related to the violation of the system's fluctuation-dissipation relation in the time-periodic NESS.
This is the Floquet generalization of the so-called Harada-Sasa relation.\cite{Harada2005} We have confirmed that this expression is indeed satisfied within the Floquet DMFT implemented with the self-consistent Migdal approximation.

 \begin{figure}[b]
  \centering
    \hspace{-0.6cm}
    \vspace{0.0cm}
   \includegraphics[width=90mm]{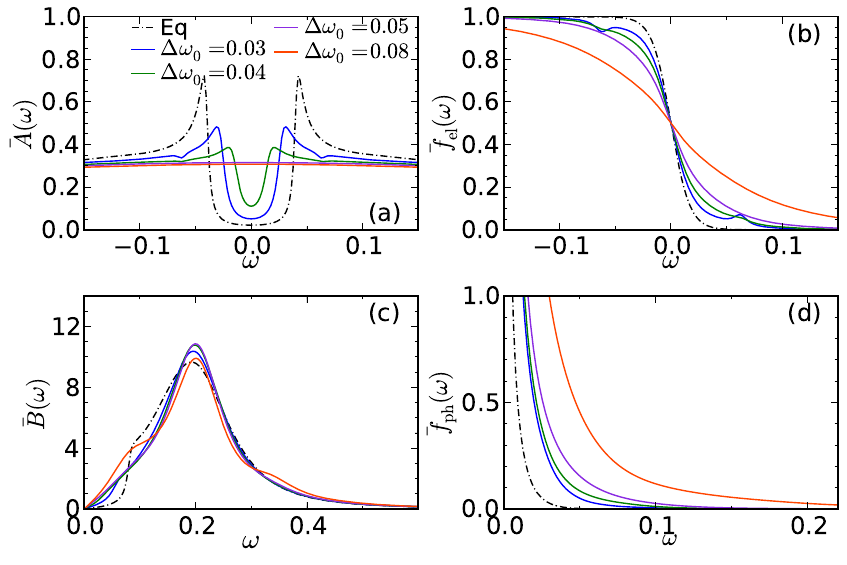} 
    \caption{Time-averaged spectra and distribution functions for the type 2 excitation for $g=0.41,\omega_0=0.4,\beta=120$ and the bath parameters $\gamma_{\rm ph}=0.02,\omega_D=0.6,\gamma_{\text {el}}=0.005,W_{\rm ebath}=2.0$ and the excitation $\Omega=0.125$. (a) Time-averaged electron spectrum. (b) Time-averaged electron distribution function. (c) Time-averaged phonon spectrum.  (d) Time-averaged phonon distribution function.}
  \label{fig:FDMFT_spectrums_Komnik}
\end{figure}



\section{Results for the type 2 excitation}
\label{appendix:type2}
In this section, we show the results for the the type 2 excitation [Eq.~(\ref{eq:Komnik})], whose effect has been discussed in Ref.~\onlinecite{Komnik2016}.
It turns out that the general behavior is very similar to the results for the type 1 excitation [Eq.~(\ref{eq:type1})] shown in the main text.
The ultimate effect of the type 2 excitation is to generally suppress the superconductivity both in the strong- and weak-coupling regimes.
This tendency becomes strong at $\Omega\simeq 2\omega_r$ and/or at $\Omega\simeq \omega_r$ as in the case of the type 1 excitation.

 \begin{figure}[t]
  \centering
      \hspace{-0.6cm}
    \vspace{0.0cm}
   \includegraphics[width=90mm]{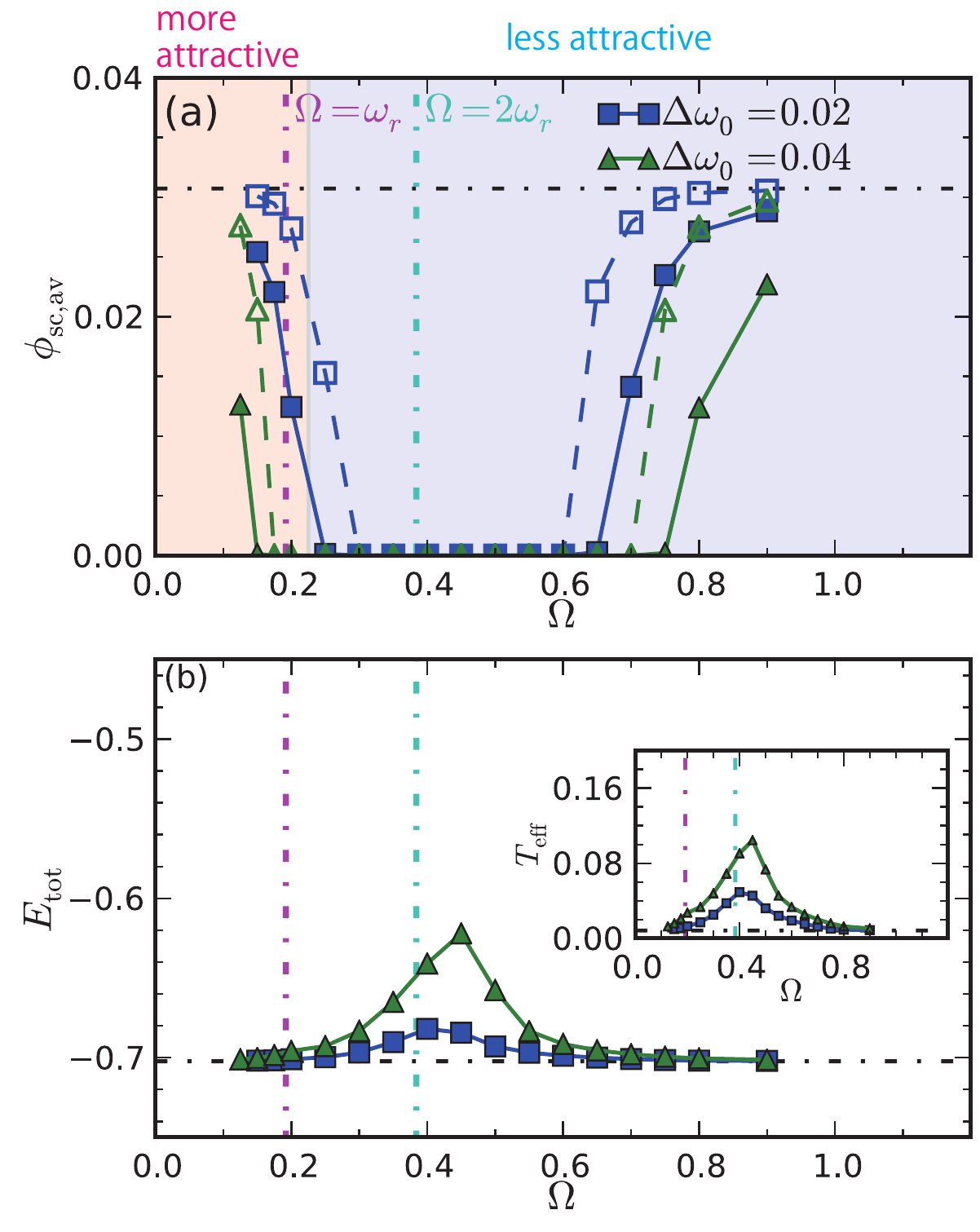} 
  \caption{Time-averaged results for the type 2 excitation for $g=0.41,\omega_0=0.4,\beta=120$ and the bath parameters $\gamma_{\rm ph}=0.02,\omega_D=0.6,\gamma_{\text {el}}=0.005,W_{\rm ebath}=2.0$. (a) Dependence of $\phi_{\rm sc,av}$ on $\Omega$ for some $\Delta \omega_0$. The open symbols show the results estimated from the effective temperature. In the red (blue) background region, $|\bar{D}^R(0)|$ increases (decreases) when $\Delta \omega_0$ changes from $0$ to $0.02$.
  (b) Dependence of $E_{\rm tot,av}$ on $\Omega$ for some $\Delta \omega_0$. The inset is the effective temperature estimated from the slope $\partial_\omega \bar{f}_{\text {el}}(\omega)|_{\omega=0}$. 
  The vertical lines indicate the position of the parametric resonance in the weakly driven limit ($\Omega=\omega_{\rm r}, 2\omega_{\rm r}$).
  } 
  \label{fig:FDMFT_summary_dum_Komnik}
\end{figure}

\subsection{Nonequilibrium steady states}

Here we show the results for the nonequilibrium steady states for the type 2 excitation in the strong electron-phonon coupling regime as in Sec.~\ref{sec:NESS}.
Even in this regime, the type 2 excitation shows an increase of the effective attractive interaction characterized by $|g^2\bar{D}^R(0)|$ at $\Omega\lesssim\omega_{\rm r}$,
and a decrease at $\Omega\gtrsim\omega_{\rm r}$.
In Fig.~\ref{fig:FDMFT_spectrums_Komnik}, we show the time-averaged spectra and distribution functions for the electrons and phonons for $\Omega=0.125$, where we observed an increase of the attractive interaction.
The general features are the same as in the case of the type 1 excitation (Fig.~\ref{fig:FDMFT_spectrums_Knap}).
The SC gap is suppressed with increasing driving amplitude and is completely wiped out at $\Delta\omega_0\ge 0.05$, see Fig.~\ref{fig:FDMFT_spectrums_Komnik}(a).
The slope of the electron distribution functions [Fig.~\ref{fig:FDMFT_spectrums_Komnik}(b)] at $\omega=0$ decreases compared to the equilibrium case, which resembles the effect of heating.
Precisely speaking, the equilibrium distribution functions at the effective temperature underestimate the number of the occupied states above the Fermi level.
The characteristic features in the NESS with SC are the kink in $\bar{A}(\omega)$ and the hump in $\bar{f}(\omega)$ at $\omega=\Omega/2$.
The renormalized phonon frequency increases as we increase the strength of the excitation,
which is naturally expected from the equilibrium state with elevated temperatures, see Fig.~\ref{fig:FDMFT_spectrums_Komnik}(c).
When we further increase the excitation strength, the Floquet sidebands of the phonons become more prominent [see $\Delta \omega=0.08$ data in Fig.~\ref{fig:FDMFT_spectrums_Komnik}(c)],
in which case the spectrum does not resemble that of the equilibrium state with elevated temperatures.

The summary of the results for the type 2 excitation is shown in Fig.~\ref{fig:FDMFT_summary_dum_Komnik}.
Generally, in the presence of the phonon driving the SC order is weakened, and this tendency becomes strong around $\Omega=2\omega_r$ as in the type 1 case, see Fig.~\ref{fig:FDMFT_summary_dum_Komnik} (a).
Again the order parameter evaluated from the effective temperature overestimates its size in the NESS, but its general behavior can be qualitatively reproduced.
The (time-averaged) total energy and the effective temperature shows a peak around $\Omega=2\omega_r$, see Fig.~\ref{fig:FDMFT_summary_dum_Komnik}(b).
The peak position is shifted to higher frequency as the excitation strength is increased, which is attributed to the hardening of the renormalized phonon frequency as in the type 1 excitation.
Quantitatively, under the type 2 excitation the SC order is more robust than under the type 1 excitation, which is consistent with the lower energy and the lower effective temperatures (compare Figs.~\ref{fig:FDMFT_summary_dum} and \ref{fig:FDMFT_summary_dum_Komnik}).

 \begin{figure}[t]
  \centering
 \hspace{-0.6cm}
    \vspace{0.0cm}
   \includegraphics[width=90mm]{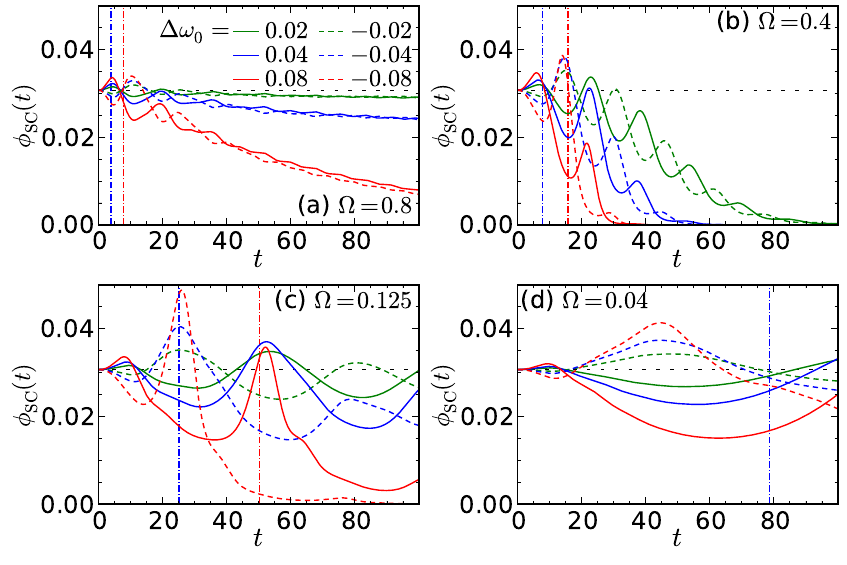}
  \caption{Time evolution of the SC order parameter under the periodic driving of type 2 for various conditions for $g=0.41,\omega_0=0.4,\beta=120$ with the bath $\gamma_{\rm ph}=0.02,\omega_D=0.6,\gamma_{\text {el}}=0.005,W_{\rm ebath}=2.0$. Horizontal black dotted lines indicate the equilibrium value, while the blue and red vertical lines show $t=\pi/\Omega$ and $t=2\pi/\Omega$.}
  \label{fig:phi_Komnik}
\end{figure}

\subsection{Transient dynamics towards the NESS}

In Fig.~\ref{fig:phi_Komnik}, we show the transient dynamics towards the NESS under the type 2 parametric phonon driving.
It corresponds to Fig.~\ref{fig:phi_knap1} in the main part.
One again finds that (i) the decease of the order parameter is fastest around $\Omega=0.4\simeq 2\omega_r$, (ii) the initial hump goes in the opposite direction 
from the adiabatic excitation, and becomes less prominent as we approach the adiabatic limit, and (iii) depending on the phase of the excitation, there can be 
an enhancement of the SC order parameter during the first cycle.
Quantitatively speaking, the speed of the melting of SC is slightly slower than in the type 1 case at the same set of the parameters.
This is consistent with the results for the NESS, where the order parameter is more robust compared to the type 1 and the effective temperature is lower than the type 1 case.
 \begin{figure}[t]
  \centering
        \hspace{-0.6cm}
      \vspace{0.cm}
   \includegraphics[width=90mm]{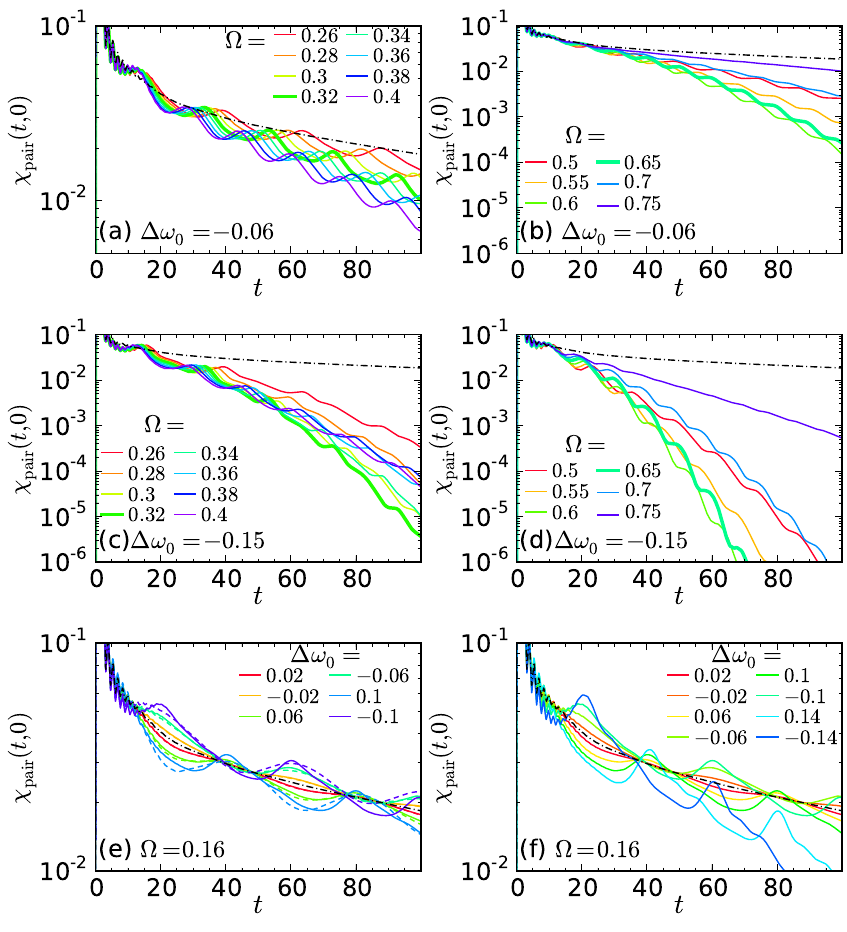}
  \caption{$\chi_{\rm pair}(t,0)$ above $T_c$ under the periodic driving of type 2. The parameters are $g=0.3,\omega_0=0.4,\beta=320$ without heat baths. 
  Lines with different colors indicate different $\Omega$. (a) (c) are for the cases around $\Omega=\omega_{\rm r}$, while (b)(d) are for the cases around $\Omega=2\omega_{\rm r}$.
  (e)(f) show the results for $\Omega=0.16$ and various $\Delta \omega_0$. 
  Dotted lines are the linear component (in $\Delta \omega_0$) estimated from the $\Delta \omega_0=\pm 0.02$ data.
  The dash-dotted black line in each panel is the equilibrium result.}
  \label{fig:sc_fluc_summary_Komnik}
\end{figure}
\subsection{Superconducting fluctuations above $T_c$}
Here we show the SC fluctuations above $T_c$ in the weak coupling regime and under the type 2 driving.
Figure~\ref{fig:sc_fluc_summary_Komnik} shows the results corresponding to Fig.~\ref{fig:sc_fluc_summary} in the main text.
First we can again see that generally the decay of the fluctuations in the driven system is faster than in the absence of phonon driving, 
which indicates that the driving does not enhance the superconductivity, see  Fig.~\ref{fig:sc_fluc_summary_Komnik} (a-d).
As in the type 1 case, around $\Omega=\omega_r$, the decay becomes particularly fastest at $\Omega=\omega_r$, even though at $\Delta \Omega=-0.06$ the tendency is not so clear.
Near $\Omega=2\omega_r$, there is again a point where the decay becomes particularly fastest. 
We also note the concave form of the $\chi_{\rm pair}(t,0)$ curve in the log scale, which is explained by the continuous heating of the system.

In Fig.~\ref{fig:sc_fluc_summary_Komnik} (e)(f), we show the results for an excitation frequency small compared to $\omega_r$. 
With the weak to moderate excitation strength, depending on the phase of $\Delta \omega_0$, $\chi_{\rm pair}(t,0)$ is shifted above or below the equilibrium curve, 
which is consistently explained by the linear component in $\Delta \omega_0$ (the dashed lines).
The non-linear component slightly but systematically shifts the SC fluctuations above the linear component, which indicates that there is more than the heating effect.
However, this slight enhancement is washed out when we further increase the excitation strength.
These features are the same as in the type 1 case.

\bibliographystyle{prsty}
\bibliography{Ref}

\end{document}